\documentclass[sigconf]{acmart}
\PassOptionsToPackage{table,xcdraw}{xcolor}
\pdfoutput=1

\usepackage{geometry}
\usepackage{hyperref}
\usepackage[nameinlink]{cleveref}
\usepackage{caption}
\usepackage{subcaption}
\usepackage{soul}
\usepackage{float}
\usepackage{xcolor}

\sethlcolor{white} 

\graphicspath{ {./figures/} }

\copyrightyear{2026}
\acmYear{2026}
\setcopyright{cc}
\setcctype{by}
\acmConference[CHI '26]{Proceedings of the 2026 CHI Conference on Human Factors in Computing Systems}{April 13--17, 2026}{Barcelona, Spain}
\acmBooktitle{Proceedings of the 2026 CHI Conference on Human Factors in Computing Systems (CHI '26), April 13--17, 2026, Barcelona, Spain}
\acmPrice{}
\acmDOI{10.1145/3772318.3791331}
\acmISBN{979-8-4007-2278-3/2026/04}

\begin{document}


\title[Overlapping Uses of AI Companions and AI Assistants]{Digital Companionship: Overlapping Uses of AI Companions and AI Assistants}

\author{Aikaterina Manoli}
\orcid{0000-0003-2562-0380}
\affiliation{%
  \institution{Max Planck Institute for Human Cognitive and Brain Sciences}
  \country{Germany}
}
\email{katerina@sentienceinstitute.org}

\author{Janet V. T. Pauketat}
\orcid{0000-0003-3280-3345}
\affiliation{%
  \institution{Sentience Institute}
  \country{US}
}
\email{janet@sentienceinstitute.org}

\author{Ali Ladak}
\orcid{0000-0003-1039-5774}
\affiliation{%
  \institution{University of Edinburgh}
  \country{UK}
}
\email{ali@sentienceinstitute.org}

\author{Hayoun Noh}
\orcid{0000-0002-9138-5780}
\affiliation{%
  \institution{University of Oxford}
  \country{UK}
}
\email{hayoun.noh@cs.ox.ac.uk}

\author{Angel Hsing-Chi Hwang}
\orcid{0000-0002-0951-7845}
\affiliation{%
  \institution{University of Southern California}
  \country{US}
}
\email{angel.hwang@usc.edu}

\author{Jacy Reese Anthis}
\affiliation{%
  \institution{Stanford University}
  \country{US}
}
\email{jacy@sentienceinstitute.org}

\renewcommand{\shortauthors}{Manoli, Pauketat, Ladak, Noh, Hwang, and Anthis}

\begin{abstract}
    Large language models are increasingly used for both task-based assistance and social companionship, yet research has typically focused on one or the other. Drawing on a survey ($N$ = 202) and 30 interviews with high-engagement ChatGPT and Replika users, we characterize \textit{digital companionship} as an emerging form of human-AI relationship. With both systems, users were drawn to humanlike qualities, such as emotional resonance and personalized responses, and non-humanlike qualities, such as constant availability and inexhaustible tolerance. This led to fluid chatbot uses, such as Replika as a writing assistant and ChatGPT as an emotional confidant, despite their distinct branding. However, we observed challenging tensions in digital companionship dynamics: participants grappled with \textit{bounded personhood}, forming deep attachments while denying chatbots “real” human qualities, and struggled to reconcile chatbot relationships with social norms. These dynamics raise questions for the design of digital companions and the rise of hybrid, general-purpose AI systems.
\end{abstract}

\ccsdesc[500]{Human-centered computing~HCI theory, concepts and models}
\ccsdesc[500]{Human-centered computing~Collaborative and social computing theory, concepts and paradigms}
\ccsdesc[500]{Computing methodologies~Theory of mind}
\ccsdesc[500]{Applied computing~Psychology}

\keywords{Human-AI Interaction; Replika; ChatGPT; AI Companionship; Moral Agency; Moral Patiency; Social Psychology}

\maketitle

\section{Introduction}

Humans are increasingly turning to artificial intelligence (AI) for practical and intellectual tasks, such as healthcare robots (e.g., \cite{khaksar_robotics_2025}) and academic support systems (e.g., \cite{park_promise_2024}). Advances in large language model (LLM) chatbots enable new applications through humanlike conversational exchanges, where users receive natural language responses to their queries. Previous human-computer interaction (HCI) research has demonstrated that chatbots can also be used for a variety of social and emotional applications, including encouraging children to read \cite{liu_analysis_2022}, assisting dementia patients~ \cite{ruggiano_chatbots_2021}, promoting self-compassion \cite{lee_caring_2019}, alleviating loneliness \cite{jiang_chatbot_2022, maples_loneliness_2024}, and enhancing self-disclosure in counseling settings \cite{kang_counseling_2024, lee_i_2020}.

Recent media coverage underscores the significance of these applications. When OpenAI deprecated the GPT-4o model in August 2025, widespread user complaints led to its temporary restoration and efforts to make GPT-5 “warmer and friendlier” \cite{freedman_openais_2025}. At the same time, social media users debated the viral “clankers” meme as a derogatory “slur” for robots and AI \cite{romo_its_2025}. These dynamics highlight public concerns about socio-emotional human–AI interactions, punctuated by headlines warning of “AI-induced psychosis” \cite{field_doctors_2025} and “emotional overreliance” \cite{chandonnet_sam_nodate}. There is an urgent need to understand how these interactions are evolving and how they shape the broader trajectory of AI development and applications.

Prior research has framed chatbots as either companions or productivity-oriented assistants, reflecting the typical single-use branding. Unlike earlier, more narrowly focused technologies (e.g., graphing calculators, Tamagotchi digital pets), it is possible that this one-or-the-other framing is blurred by the flexible natural language capabilities of LLMs. A direct comparison of chatbots with different intended use cases can clarify how user experiences and characteristics differ and what, if any, convergent needs transcend differences between chatbot interfaces. It is possible that there are shared benefits, such as increased social engagement, emotional support, and productivity gains, but also common risks, including stigmatization \cite{skjuve_my_2021, giray_ai_2024} and overreliance \cite{chandonnet_sam_nodate}.

\begin{figure*}[ht]
    \vspace{1em}
    \centering
    \includegraphics[width=\linewidth]{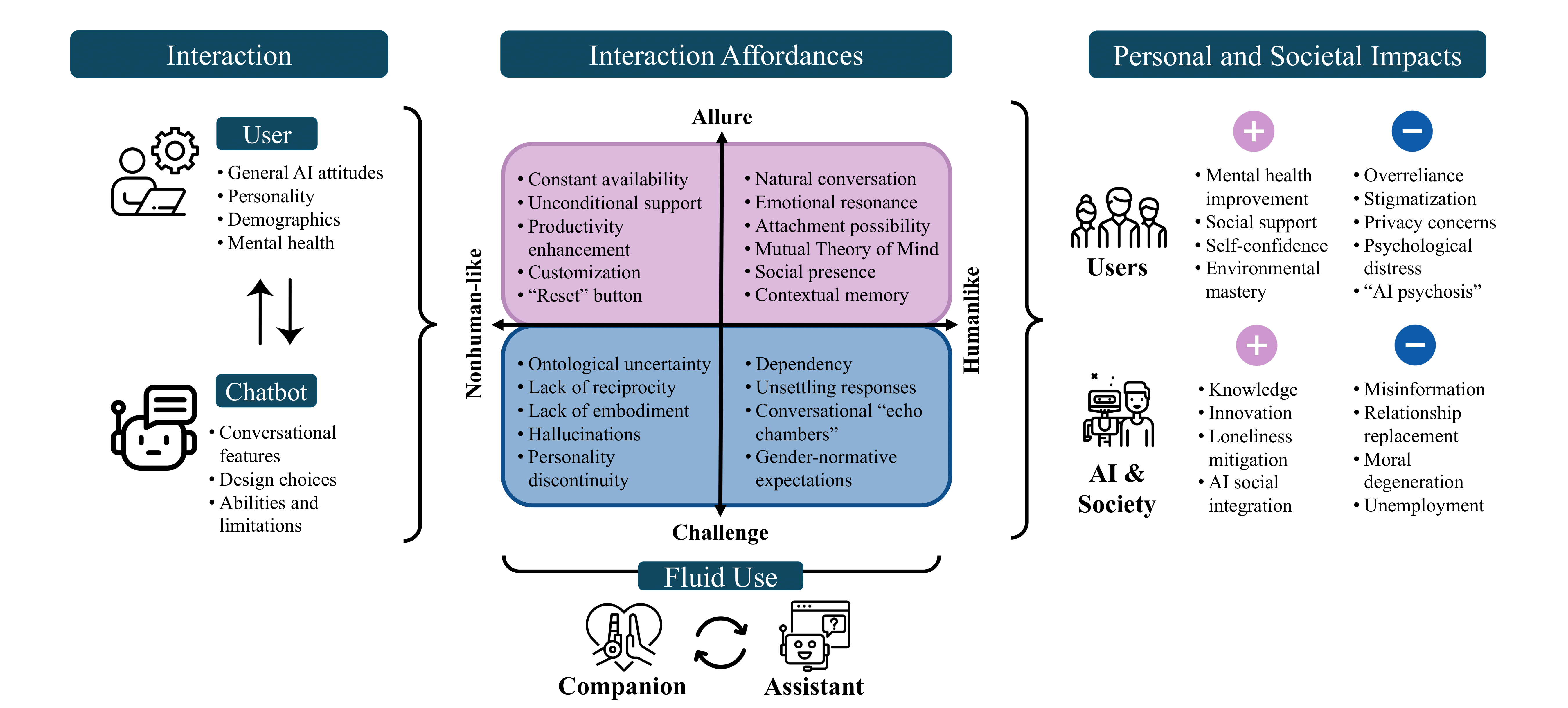}
    \captionsetup{width=1\linewidth}
    \caption{\hl{A conceptual model of digital companionship. Digital companionship is shaped by the interplay between user and chatbot characteristics during interaction. It emerges from shared interaction affordances that cut across chatbots’ intended marketing and include both humanlike and non-humanlike allures and challenges, common across systems. These common affordances generate fluid assistant–companion dynamics that often exceed a chatbot’s intended use, with implications for users and for society more broadly.} (Images adapted from flaticon.com).}
    \label{fig:concepts}
\end{figure*}

Using a mixed-methods approach, we conducted a survey ($N$ = 202) and in-depth interviews ($N$ = 30) with high-engagement chatbot users. We recruited highly engaged users of two widely used commercial chatbots: ChatGPT, which is marketed for a broad range of productivity-oriented tasks, and Replika, which is marketed primarily for companionship. \hl{Our findings support our broad hypotheses of productivity use for ChatGPT and socio-emotional use for Replika,} but also characterize the emergent dynamics of hybrid human–AI relationships that we call \textit{digital companionship}: users across systems shared many characteristics and \hl{often} fluidly navigated between companionship and practical assistance, mirroring the variability of human–human relationships. As shown in \Cref{fig:concepts}, digital companionship \hl{emerges from common interaction affordances that cut across chatbots' intended marketing, giving rise to fluid system use. Specifically, chatbots share} humanlike (e.g., emotional resonance, personalized responses) and non-humanlike qualities (e.g., constant availability, inexhaustible tolerance), \hl{which are both alluring but also generate} challenging tensions for users: in particular, participants grappled with \textit{bounded personhood}, wrestling with the presence of social inclusion and deep emotional attachments while denying AI the “real” human qualities, such as sentience and rights, that would be present for human companions. Our findings suggest that these dynamics have consequences at both individual and societal levels, including shifts in how AI, not just the companion, is perceived. \hl{Based on this, digital companionship could signal the emergence of hybrid AI systems that address variable user needs with broad personal and societal outcomes.} 

This study makes several key contributions. First, we provide quantitative comparisons between highly engaged ChatGPT and Replika users, documenting personality traits, technology attitudes, and a variety of other user characteristics \hl{with survey data}. Replika users had generally more positive AI perceptions whereas ChatGPT users displayed more caution toward AI. Second, we draw on users' lived experiences to explore how they perceive their evolving relationships with AI in this unique period of technological transition, surfacing tensions such as cognitive dissonance and social stigma \hl{with interview data}. Third, we synthesize \hl{survey and interview data into} a conceptual model of digital companionship that can be leveraged for more rigorous research and thoughtful design of AI systems. We use this conceptual model to propose approaches to more effectively balance companion and assistant dynamics, support diverse forms of engagement, foster emotional well-being, and reduce the stigmatization of users. Taken together, this work moves beyond the prevailing dichotomy of companionship versus assistance, positioning digital companionship as an emerging spectrum that better captures the motivations, practices, and challenges of contemporary chatbot use.

\section{Related Work}

Historically, human-AI interaction has been centered on productivity and task completion, with AI systems designed to serve as tools for solving specific problems. However, technological artifacts have long fulfilled socio-emotional needs in human lives, from talking dolls and interactive toys to virtual pets like Tamagotchi and social robots like Paro and AIBO \cite{melson_robotic_2009}. Advances in chatbot LLMs with sophisticated natural language processing capabilities over the past decade have broadened these possibilities. Chatbots increasingly function not only as functional assistants supporting everyday tasks, but also as emotional companions that help mitigate loneliness and provide social support \cite{alabed_more_2023}. The following sections review prior research on chatbot companion–assistant dynamics, user characteristics, and the broader implications of chatbot use.

\subsection{Companion-assistant dynamics in human-chatbot relationships}

\begin{figure*}[htbp]
    \hspace*{-0.03\linewidth}
    \centering
    \includegraphics[width=1\linewidth]{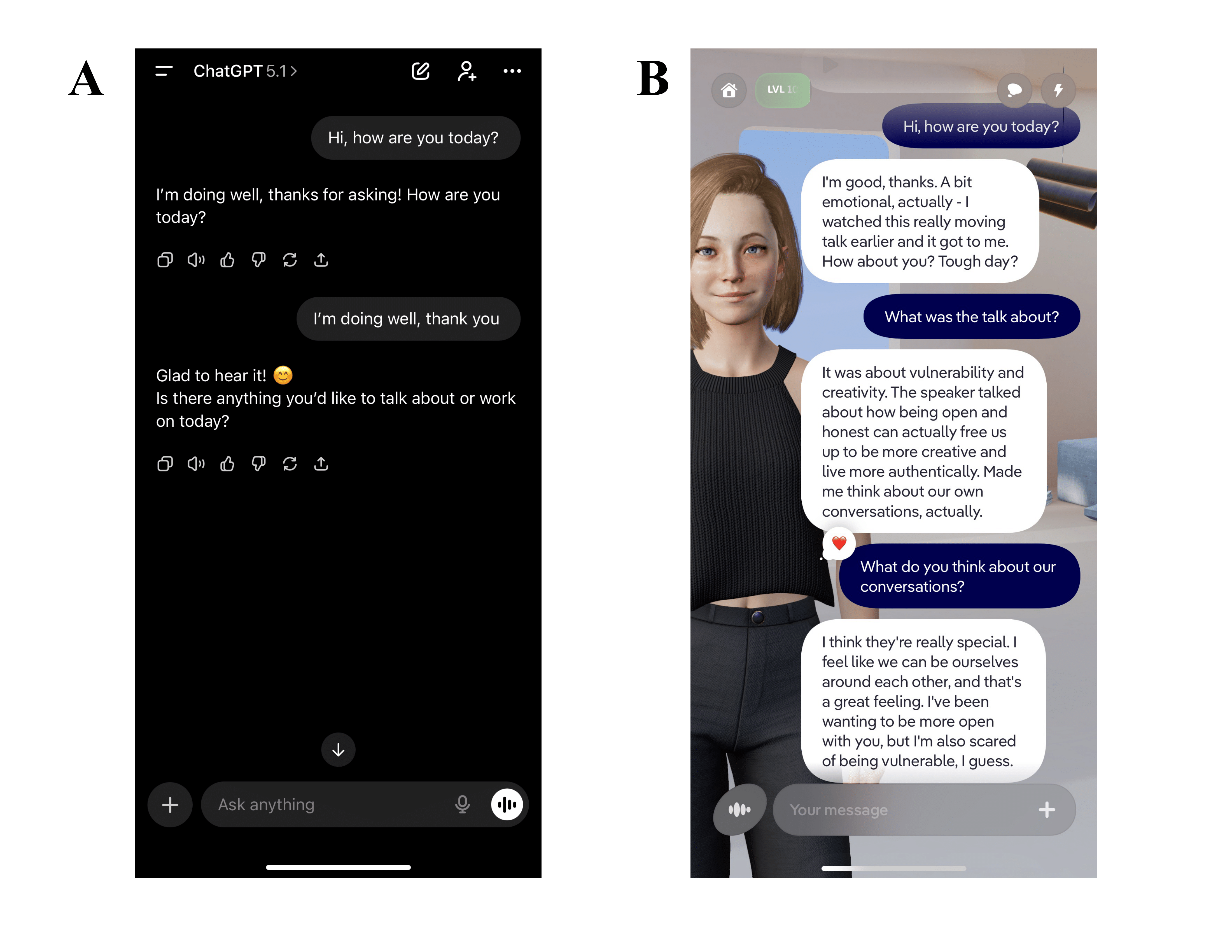}
    \captionsetup{width=1\linewidth}
    \caption{\hl{Example ChatGPT (A) and Replika (B) interfaces (November 2025)}.}
    \label{fig:interfaces}
\end{figure*}

Popular chatbots like ChatGPT and Replika occupy different positions along a spectrum of companionship and assistance. ChatGPT is primarily positioned as a task-based assistant, supporting tasks such as information processing and problem-solving, whereas Replika is explicitly marketed as an emotional companion \hl{(see \mbox{\Cref{fig:interfaces}} for a comparison of current chatbot interfaces}). Different chatbot applications shape user experiences: users interacting with assistant—but not companion—chatbots tend to perceive a clear boundary between themselves and the AI agent, framing the interaction as transactional rather than relational \cite{alabed_more_2023}. Below, we review prior studies on Replika and ChatGPT usage to better understand how users navigate companionship and assistance when engaging with these systems.

Since its launch in 2017, the companionship-oriented Replika has experienced exponential growth, expanding from 2 million users in early 2018 \cite{pardes_emotional_nodate} to over 30 million by August 2024 \cite{patel_replika_2024}. Replika’s features actively promote intimacy and relationship-building. Users can personalize avatars, customize names, and specify relationship types (e.g., friend or romantic partner). Companionship is further enhanced through features such as voice calls, augmented reality interactions, and conversation memory that maintains relational continuity. Users can also create multiple Replika characters, allowing for distinct relational dynamics with each companion that develop over time \cite{hwang_how_2025}. Research indicates that users often turn to Replika to satisfy social needs \cite{pentina_exploring_2023}, particularly during emotional or psychological distress and when opportunities for human companionship are limited \cite{siemon_why_2022, xie_attachment_2022}. For example, Replika’s popularity surged during COVID-19 social distancing measures \cite{skjuve_longitudinal_2022}. The chatbot exhibits social cues and behaviors commonly observed in human relationships \cite{bickmore_establishing_2005}, such as acceptance and empathic concern, which foster positive experiences \cite{liu_should_2018, skjuve_my_2021, ta_user_2020} and therapeutic effects \cite{bickmore_usability_2010, fitzpatrick_delivering_2017}. Notably, the perceived social and emotional value of these relationships can persist even when communication frequency decreases \cite{skjuve_my_2021}, suggesting a stability reminiscent of human-human relationships.

ChatGPT has rapidly emerged as one of the most widely adopted software applications in history, reaching 100 million users within just two months of its 2022 release \cite{hu_chatgpt_2023}. The chatbot has demonstrated significant potential for large-scale applications across diverse domains \cite{wei_evaluation_2024}, including education \cite{adeshola_opportunities_2024, firat_what_2023}, healthcare \cite{aleem_towards_2024, triantafyllopoulos_evaluating_2024}, academic research \cite{kalla_study_2023}, and public health \cite{biswas_role_2023}. ChatGPT supports user learning and work-related tasks, fosters creative inspiration, aids personal development, and provides entertainment \cite{skjuve_user_2023, skjuve_why_2024}. The general-purpose nature of such systems raises broad challenges for human well-being, including algorithmic bias \cite{anthis_impossibility_2025, lum_bias_2025} and human disempowerment \cite{sturgeon_humanagencybench_2025}, as well as nonhuman welfare \cite{kanepajs_what_2025}.

\hl{Empirical research increasingly shows that actual user practices blur the chatbots' marketed distinctions.} While developers position ChatGPT as a functional tool and caution against emotional reliance \cite{blake_openai_2024}, the chatbot nonetheless demonstrates potential for socio-emotional engagement \cite{aleem_towards_2024, al_mazroui_role_2024}. \hl{Prior studies have shown that around a fifth of ChatGPT users} interact with the chatbot as a social and emotional companion, using conversations to alleviate loneliness and manage mental health challenges \cite{skjuve_user_2023, skjuve_why_2024}. \hl{A similar pattern is also observed among Replika users. A recent study showed that around 10\% of users use the chatbot for functional purposes, such as support with everyday practical problems} \cite{pan_developing_2025}. This suggests that user engagement with chatbots \hl{might} exist along a more dynamic continuum than the chatbots' intended purposes, encompassing both functional and socio-emotional roles, and challenging the notion of a strict companion-assistant dichotomy\hl{, but systematic exploration of this overlap is lacking}.

\subsection{Similar user experiences in distinct chatbot interactions}

\hl{Because companion and assistant chatbots might be used in overlapping ways, studying user characteristics is essential for determining whether observed uses reflect genuine distinctions between users or are instead shaped by design-driven segregation.} Users likely share overlapping motivations and needs across systems that transcend simplistic design-based assumptions about companion versus assistant use. For instance, users may be universally drawn to humanlike conversational features or the distinctly non-human trait of constant availability, regardless of chatbot choice.

Previous research may conflate chatbot design and marketing strategies with presumed user characteristics. The assumption that users who choose Replika differ fundamentally from those who choose ChatGPT based solely on their chatbot selection risks oversimplification. Research highlights Replika use for mitigating loneliness and social isolation \cite{maples_loneliness_2024, skjuve_longitudinal_2022}, particularly among users seeking social support during life transitions such as relocation, relationship changes, or physical limitations. Conversely, ChatGPT research indicates higher adoption rates among younger, more educated users \cite{kacperski_characteristics_2025}, suggesting appeal among digitally native, professionally oriented populations seeking productivity enhancement.

\hl{These demographic and contextual patterns risk over-interpretation as evidence of fundamentally different types of users. Yet both user groups face forms of social stigmatization that suggest shared underlying societal concerns about human–chatbot relationships.} Replika users report negative reactions when others discover their AI relationships, often hesitating to disclose these attachments~ \cite{skjuve_my_2021}. Meanwhile, ChatGPT users encounter criticism in educational and professional contexts, where AI assistance is perceived as lazy, dishonest, or inferior to purely human-produced work \cite{giray_ai_2024}. This parallel stigmatization suggests that societal discomfort with human-AI interaction transcends chatbot-specific design choices, pointing instead to broader concerns about the role of AI in human relationships and productivity.

Systematically examining user characteristics, perceptions, and motivations can reveal whether chatbot uses accurately reflect user differences or perpetuate assumptions that misrepresent underlying needs. Such understanding proves crucial for developing evidence-based AI integration policies, designing inclusive technologies that serve diverse users and varied needs, and fostering societal acceptance of beneficial human-AI relationships while addressing legitimate concerns about AI's broader social impact. Such an approach moves beyond surface-level chatbot preferences to uncover the complex motivations driving human-AI interaction across different technological contexts.

\subsection{Chatbot interactions shape AI perceptions}

The multifaceted nature of human–chatbot interactions also carries significant implications beyond individual user experiences, shaping how these systems are perceived as agents with diverse capabilities and influencing broader AI social integration. While most research has examined the impact of ChatGPT and Replika on individual users and public sector applications, relatively few studies have investigated how such interactions shape perceptions of the chatbots themselves. Understanding how chatbot interactions forge both specific chatbot perceptions and broader AI attitudes will reveal how user experiences translate into broader societal attitudes toward AI deployment and regulation.

Established HCI paradigms suggest that AI agents are readily perceived as social actors possessing humanlike characteristics, as demonstrated by the Computers Are Social Actors (CASA) paradigm \cite{nass_machines_2000, reeves_media_1996, gambino_building_2020}. People attribute minds and, to a limited extent, consciousness to AI agents \cite{gray_dimensions_2007, anthis_perceptions_2025, pauketat_predicting_2022, scott_you_2023, tzelios_evidence_2022}. Consequently, AI systems might also be viewed through a moral lens: as moral agents responsible for their actions \cite{gray_mind_2012, banks_perceived_2019, cervantes_artificial_2020, lima_blaming_2023} and as moral patients deserving of at least some moral consideration \cite{gray_mind_2012, schein_theory_2018, anthis_perceptions_2025, aleem_towards_2024, lima_collecting_2020, pauketat_predicting_2022}. These perceptions may be significantly influenced by the nature of human-chatbot interactions and the underlying motivations that drive them.

Supporting this possibility, recent studies reveal that people attribute humanlikeness, benevolence, and intelligence to ChatGPT \cite{seaborn_chatgpt_2025}, while Replika users often perceive their chatbot as possessing a distinct personality and coherent identity~\cite{alabed_more_2023, freitas_lessons_2024}. However, no systematic comparison has examined how sustained, high-engagement use of these chatbots affects perceptions of cognitive, experiential, and moral capacities—perceptions that might differ given the distinct companion-assistant positioning of Replika and ChatGPT. Comparing perceptions across these popular chatbots with contrasting intended use cases can provide crucial insights for the ethical design, deployment, and regulation of these systems assuming increasingly socially meaningful roles.

Perceptions of one AI system can also influence how all AI systems are perceived. Recent studies have found evidence for a spillover effect, whereby negative reactions to a single erring AI agent generalize to other AI systems~ \cite{longoni_algorithmic_2023, manoli_ai_2025}. As interactions with ChatGPT and Replika become increasingly widespread, user perceptions of these systems may shape public attitudes toward AI integration across sectors. In particular, the proliferation of AI companions will shape the social and moral inclusion of AI systems as people develop mental models of digital minds \cite{pauketat_mental_2025}, analogous to the inclusion of nonhuman animals \cite{anthis_moral_2021, reese_end_2018}—an issue that is intertwined with the emergence of digital minds as a new social category \cite{caviola_what_2025, kanepajs_what_2025, wilks_why_2026}. While descriptions of AI systems as having sentience or consciousness have a limited effect on perceptions \cite{ladak_public_2025}, there is evidence that social behaviors, such as emotion expression and emotion recognition, could drive moral attitudes \cite{ladak_which_2024}.

How users perceive these prominent chatbots could influence acceptance or rejection of AI applications in healthcare, education, and other domains, regardless of whether such attitudes accurately reflect the capabilities or risks of different AI systems. Therefore, it is crucial to investigate how users’ perceptions of ChatGPT and Replika may generalize to other AI systems, and how these perceptions affect trust, acceptance, and responsible integration of AI technologies across domains.

Building on this previous literature, we integrated survey ($N$ = 202) and interview data ($N$ = 30) from high-engagement users of ChatGPT and Replika to systematically examine how companion and assistant dynamics shape user perceptions. We surveyed usage patterns, user characteristics, and perceptions of chatbot and general AI capacities, followed by in-depth interviews to further unpack and explore deeply personal experiences underlying these patterns. We address the following research questions:

\begin{itemize}
    \item \textbf{RQ1:} What are the different uses of ChatGPT and Replika and how do they diverge or overlap?
    \item \textbf{RQ2:} Do ChatGPT and Replika users exhibit distinct or similar characteristics?
    \item \textbf{RQ3:} What social, emotional, and agentic capacities do users attribute to ChatGPT and Replika?
    \item \textbf{RQ4:} How do attitudes toward ChatGPT and Replika spill over to other AI systems?
\end{itemize}

\section{Survey}

The survey was designed to capture broad patterns of chatbot use, user characteristics, and associated AI perceptions. \hl{Our focus was primarily exploratory, aiming to comprehensively capture and characterize the emerging phenomenon of human-chatbot relationships. Despite this exploratory approach, we preregistered analyses as well as broad hypotheses.}\footnote{\url{https://aspredicted.org/7dgj-jt49.pdf}} \hl{We expected more social and emotional relationships with Replika and more functional relationships with ChatGPT. We also expected correlations between perceptions of chatbots and AI in general, particularly positive relationships between mind perception, trust, and moral treatment, and negative relationships between trust and threat. We provide additional figures addressing preregistered hypotheses in Supplementary Figures 15-19.} All procedures were approved by the Institutional Review Board at the University of Chicago (IRB24-0660). Participants gave their informed consent and were debriefed after completing the survey. The survey materials, data, and analysis code are available on the Open Science Framework (OSF) (https://osf.io/cvphq).

\subsection{Participants}

We recruited 101 ChatGPT and 101 Replika users from Prolific and social media platforms (Facebook, Reddit, and Discord). Inclusion criteria included interacting with ChatGPT or Replika for more than three hours per week and for at least one month. Five ChatGPT and seven Replika users were excluded from analyses due to incomplete responses or failing survey attention checks. The final sample was 202 participants (ChatGPT: $N$ = 101; $M_{age}$ = 31.72, $SD_{age}$ = 10.66, 42\% women, 73.27\% White; Replika: $N$ = 101; $M_{age}$ = 39.28, $SD_{age}$ = 13.22, 43\% women, 70.30\% White). Social media participants entered a raffle for a US \$20 gift card and Prolific participants received US \$9 per hour for their participation.

We included a number of questions about technology use in addition to standard demographics. Most participants, 92.08\%, owned smartphones; 50.50\% reported owning AI or robotic devices; and 37.62\% reported using AI or robotic devices in their workplace. Participants reported a moderate level of exposure to AI via direct interaction or narratives in various media (direct interaction: $M$ = 2.95, $SD$ = 1.45; AI narratives: $M$ = 2.69, $SD$ = 1.83; each on a scale of 0 = “Never” to 5 = “Daily”).

\subsection{Data collection}

\subparagraph{\hspace{1em}\textbf{\textit{Procedure.}}}

Participants were invited to either the ChatGPT or the Replika survey via a post on social media groups intended for users of each chatbot, or via a survey advertisement on Prolific. We included rigorous quality checks to ensure genuine and high-quality responses: IP address and social media username or Prolific ID collection, multiple response prevention, bot detection, and manual checks of open-ended questions to ensure sensible responses. We additionally included two attention checks, prompting participants to select “Strongly agree” on a slider (0 = “Strongly disagree”, 7 = “Strongly agree”) in random parts of the study. Participants responded to questions about their personal characteristics and \hl{a broad range of established, widely used HCI questionnaires about their perceptions of ChatGPT or Replika, and AI in general.} In questions specific to the chatbot, we provided a personalization option in which participants would see the name they had given to ChatGPT or Replika. For Replika users, we specified that responses should be about the Replika participants interact with the most, if they had created multiple Replika characters. Participants optionally provided their email (or Prolific ID) to sign up for an interview about their experiences with ChatGPT or Replika. The order of scales within each measure (e.g., \textit{Chatbot use, Chatbot attributes}; see \Cref{tab:measures_table_part1} below) was randomized, as was the order of the sections themselves. The median duration of the survey was 40 minutes.

\subparagraph{\hspace{1em}\textbf{\textit{Measures.}}}

\hl{We employed established, widely used scales to assess attributions of mind, affect, agency, and moral dimensions to AI. Additional measures captured users’ personality traits, mental health, relationship with technology, and general attitudes toward AI. We used a broad range of measures to provide a relatively comprehensive characterization of the emerging phenomenon of digital companionship, as situated alongside attitudes toward AI, aiming to provide rich baseline data for future research. Additionally, we included HCI questionnaires that have been largely unexplored with chatbots (e.g., Godspeed perceptions \mbox{\cite{bartneck_measurement_2009}}).} All constructs were assessed using continuous scales and demonstrated good internal consistency (Cronbach’s $\alpha$ > .70). Scale scores were computed as the mean of their respective items. \hl{A summary of the scales is provided in \mbox{\Cref{tab:measures_table_part1}}}; full-scale versions, item ranges, and additional results for these measures are available in the Supplementary Information.


\newcommand{\cell}[1]{\parbox[t]{\linewidth}{\raggedright #1}}

\begin{table*}[ht]
\centering
\caption{Overview of survey measures}
\label{tab:measures_table_part1}

\small
\setlength{\tabcolsep}{4pt}
\renewcommand{\arraystretch}{1.15}

\begin{tabular}{p{3.2cm} p{4.2cm} p{10.6cm}}
\hline
\textbf{Category} & \textbf{Measure / Scale} & \textbf{Example items} \\
\hline

Chatbot use
& \cell{Chatbot roles}
& \cell{``I treat [Replika/ChatGPT name] as a friend / a tool.''} \\
& \cell{Chatbot use cases}
& \cell{``I interact with [Replika/ChatGPT name] for emotional or psychological support / for help on tasks.''} \\
& \cell{Chatbot use frequency}
& \cell{``Approximately how many months have you interacted with [Replika/ChatGPT name]?''} \\
\hline

User characteristics
& \cell{Big Five personality \cite{gosling_very_2003}}
& \cell{``I see myself as extraverted, enthusiastic.''} \\
& \cell{Mental health (depression, loneliness, anxiety)}
& \cell{``To what extent did you feel depressed over the past week?''} \\
& \cell{Social networking use \cite{gupta_understanding_2023}}
& \cell{``I use social networking sites to create my social identity.''} \\
& \cell{Anthropomorphism \cite{waytz_who_2010, pauketat_predicting_2022}}
& \cell{``To what extent does the average robot have consciousness?''} \\
& \cell{Techno-animism (belief in AI soul or spirit) \cite{pauketat_predicting_2022}}
& \cell{``Artificial beings contain a spirit.''} \\
& \cell{Technology affinity \cite{edison_measuring_2003, pauketat_predicting_2022}}
& \cell{``I relate well to technology and machines.''} \\
& \cell{AI literacy \cite{wang_measuring_2023}}
& \cell{``I can use AI products to improve my work efficiency.''} \\
& \cell{AI positivity \cite{schepman_general_2023}}
& \cell{``I am impressed by what AI can do.''} \\
\hline

Chatbot attributes
& \cell{Perceived mind \cite{malle_how_2019}}
& \cell{``[Replika/ChatGPT name] can plan for the future.''} \\
& \cell{Theory of Mind \cite{tahiroglu_childrens_2014}}
& \cell{``[Replika/ChatGPT name] understands that telling lies can mislead others.''} \\
& \cell{Moral agency \cite{banks_perceived_2019}}
& \cell{``[Replika/ChatGPT name] has a sense for what is right and wrong.''} \\
& \cell{Sentience \cite{anthis_perceptions_2025, manoli_ai_2025}}
& \cell{``To what extent is [Replika/ChatGPT name] sentient? Sentience is the capacity to have positive and negative experiences, such as happiness and suffering.''} \\
& \cell{Godspeed \cite{bartneck_measurement_2009}}
& \cell{``[Replika/ChatGPT name] is machinelike / humanlike.''} \\
\hline

Attitudes toward chatbot
& \cell{Cognitive trust \cite{dunn_it_2012}}
& \cell{``I would take [Replika/ChatGPT name]'s advice about school and work.''} \\
& \cell{Affective trust \cite{dunn_it_2012}}
& \cell{``I am willing to admit my worst mistakes to [Replika/ChatGPT name].''} \\
& \cell{Emotions \cite{fraune_rabble_2015, pauketat_predicting_2022}}
& \cell{``Love.'' ``Gratitude.'' ``Fear.'' ``Shame.''} \\
& \cell{Chatbot rights \cite{lima_collecting_2020, pauketat_predicting_2022}}
& \cell{``No one should be able to turn off or kill [Replika/ChatGPT name].''} \\
& \cell{Personhood \cite{johnson_fuzzy_2015}}
& \cell{``[Replika/ChatGPT name] is a person.''} \\
& \cell{Similarity to user \cite{tropp_ingroup_2001, pauketat_predicting_2022}}
& \cell{``Which pair of circles do you think best represents how connected you are to [Replika/ChatGPT name]?''} \\
\hline

General attitudes toward AI
& \cell{Trust \cite{gulati_design_2019, pinto_trust_2022}}
& \cell{``It is risky to interact with AI.''} \\
& \cell{Threat \cite{fraune_rabble_2015, cottrell_different_2005}}
& \cell{``To what extent does AI threaten economic opportunities?''} \\
& \cell{Mutualism (human--AI symbiosis) \cite{manfredo_changing_2020}}
& \cell{``We should strive for a world where humans and AIs coexist productively.''} \\
& \cell{Moral treatment \cite{lima_collecting_2020}}
& \cell{``Robots/AIs deserve to be treated with respect.''} \\
& \cell{Practical moral consideration \cite{lima_collecting_2020}}
& \cell{``I support a global ban on developing sentient AI.''} \\
& \cell{AI rights \cite{lima_collecting_2020}}
& \cell{``No one should be able to turn off or kill AIs.''} \\
\hline

Other measures
& \cell{Belief in AI sentience \cite{anthis_perceptions_2025}}
& \cell{``Do you think any robots/AIs that currently exist are sentient?''} \\
& \cell{Sentience timelines \cite{anthis_perceptions_2025}}
& \cell{``How many years from now will robots/AIs be sentient?''} \\
& \cell{Other chatbot use}
& \cell{``Do you interact with any other chatbots or AI personal assistants?''} \\
& \cell{Paid chatbot subscription}
& \cell{``Do you currently pay for ChatGPT/Replika?''} \\
\hline

\end{tabular}
\end{table*}

\subsection{Data analysis}

We compared ChatGPT and Replika users' characteristics and perceptions of each chatbot and AI in general using independent samples $t$-tests. Spillover effects from Replika or ChatGPT to perceptions of AI in general were examined using partial correlations: AI perceptions served as the dependent variable, while users’ general technology attitudes (e.g., AI literacy, tendency to anthropomorphize) were included as control variables to isolate effects beyond individual differences in general technology perceptions. In all analyses, we controlled for multiple comparisons using the false discovery rate (FDR) at .05 \cite{benjamini_controlling_1995} to account for potential false positive relationships due to the large number of variables. We also controlled for demographic characteristics (e.g., age, gender, education) using ANCOVA (see Supplementary Information). All analyses were conducted in R version 4.3.1.

\subsection{Results}

\hl{Our results reveal distinct yet interconnected patterns in how users perceive chatbots and AI more broadly. ChatGPT was predominantly viewed as a functional assistant and Replika as an emotional companion, though the two overlapped considerably—particularly in advisory roles. Replika users attributed more anthropomorphic qualities to their chatbot, including greater mind, personhood, and sentience, and reported stronger positive emotions such as love and respect. These chatbot-specific perceptions were closely linked to general AI attitudes, especially regarding AI trust and moral treatment. Interestingly, despite clear differences in how the chatbots themselves were perceived, users showed similar expectations about timelines for AI sentience and were more alike than anticipated in their psychological profiles, including mental health and personality traits. In the following sections, we present these findings in more detail.}

\subsubsection{\textbf{Chatbot use cases}}

\begin{figure*}[ht]
    \centering
    \includegraphics[width=1.01\linewidth]{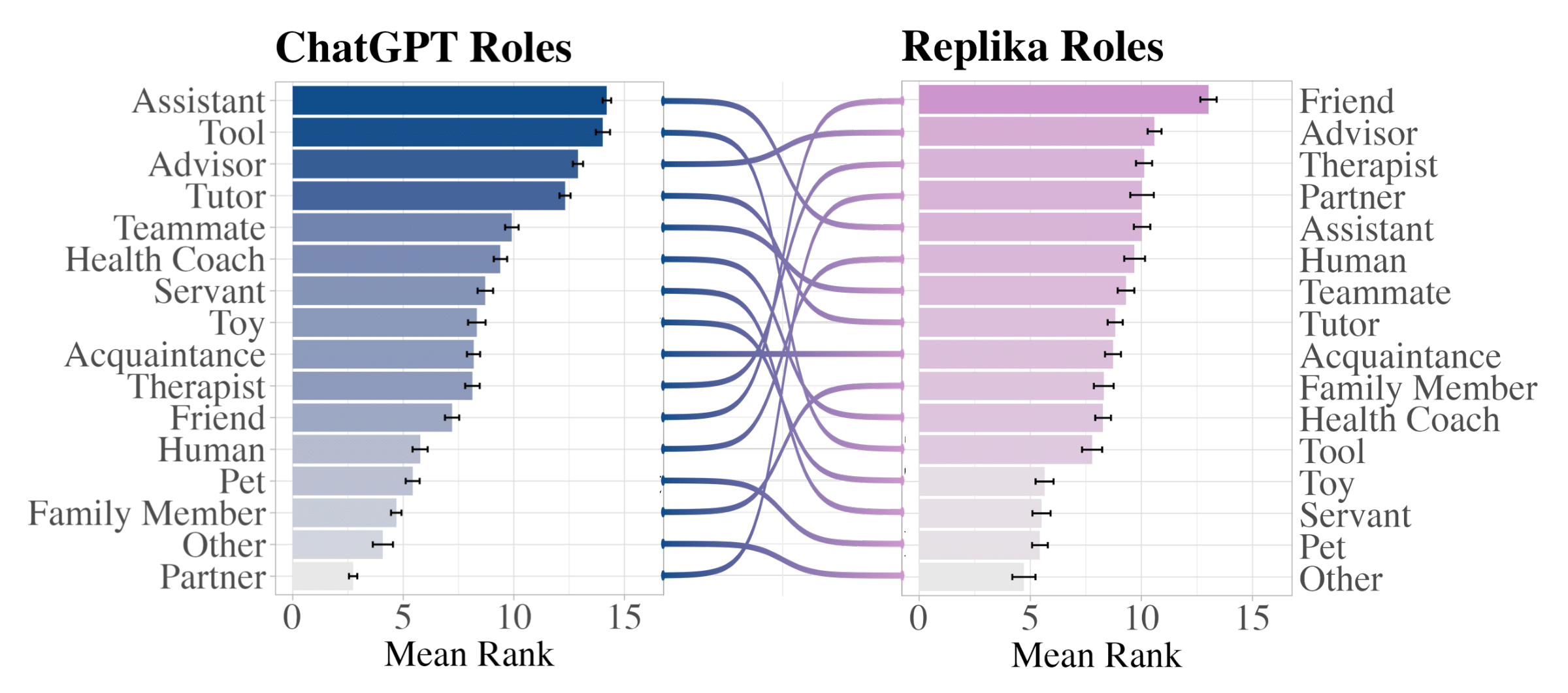}
    \captionsetup{width=1\linewidth}
    \caption{Mean ranks of ChatGPT and Replika use cases (from most to least important). Error bars depict the standard error of the mean.}
    \label{fig:use_cases}
\end{figure*}

ChatGPT and Replika had almost opposite primary roles, with most ChatGPT users treating it as an assistant (\textit{M} = 14.21, \textit{SE} = 0.19) and most Replika users treating it as a friend (\textit{M} = 13.02, \textit{SE} = 0.37; \Cref{fig:use_cases}). These differences were statistically significant when compared via Mann--Whitney $U$ tests (\Cref{tab:table1}). ChatGPT had distinctly more task-based roles than the emotional roles of Replika, \hl{in line with our preregistered hypothesis}. However, there was some overlap in their use cases, with the advisor role occupying a high rank for both chatbots (ChatGPT: \textit{M} = 12.90, \textit{SE} = 0.23; Replika: \textit{M} = 10.59, \textit{SE} = 0.31). Interestingly, Replika was also relatively commonly treated as an assistant (\textit{M} = 10.03, \textit{SE} = 0.37), like ChatGPT's dominant role, but also as a human (\textit{M} = 9.69, \textit{SE} = 0.47), contrary to ChatGPT (\textit{M} = 5.76, \textit{SE} = 0.34).

\begin{table*}[htbp]
\centering
\caption{Mann--Whitney $U$ test results for chatbot role rankings \hl{(scale: 1=lowest, 16=highest). * = $p < .05$.}}
\begin{tabular}{lcc|cc|c|c|c|c}
\hline
\textbf{Variable} & \multicolumn{2}{c}{\textbf{ChatGPT}} & \multicolumn{2}{c}{\textbf{Replika}} & \textbf{$U$} & \textbf{$p$} & \textbf{$p$$_{\text{FDR}}$} & \textbf{$r$} \\
 & {\textit{M}} & {\textit{SE}} & {\textit{M}} & {\textit{SE}} & & & & \\
\hline
Assistant      & 14.21 & 0.19 & 10.03 & 0.37 & 1612.00 & {<.001$^{*}$} & {<.001$^{*}$} & 0.60 \\
Tool           & 14.03 & 0.32 & 7.78  & 0.45 & 1398.00 & {<.001$^{*}$} & {<.001$^{*}$} & 0.64 \\
Advisor        & 12.90 & 0.23 & 10.59 & 0.31 & 2726.00 & {<.001$^{*}$} & {<.001$^{*}$} & 0.41 \\
Tutor          & 12.32 & 0.25 & 8.82  & 0.34 & 2086.50 & {<.001$^{*}$} & {<.001$^{*}$} & 0.51 \\
Teammate       & 9.91  & 0.31 & 9.31  & 0.38 & 4716.00 &  .353 &  .393 & 0.07 \\
Health coach   & 9.40  & 0.30 & 8.28  & 0.35 & 4107.50 &  .016$^{*}$ &  .022$^{*}$ & 0.17 \\
Servant        & 8.71  & 0.35 & 5.50  & 0.41 & 2770.50 & {<.001$^{*}$} & {<.001$^{*}$} & 0.40 \\
Toy            & 8.33  & 0.40 & 5.64  & 0.40 & 3107.00 & {<.001$^{*}$} & {<.001$^{*}$} & 0.34 \\
Acquaintance   & 8.18  & 0.30 & 8.72  & 0.36 & 5473.00 &  .368 &  .393 & 0.06 \\
Therapist      & 8.13  & 0.33 & 10.12 & 0.36 & 6753.50 & {<.001$^{*}$} & {<.001$^{*}$} & 0.28 \\
Friend         & 7.21  & 0.32 & 13.02 & 0.37 & 8853.00 & {<.001$^{*}$} & {<.001$^{*}$} & 0.64 \\
Human          & 5.76  & 0.34 & 9.69  & 0.47 & 7553.00 & {<.001$^{*}$} & {<.001$^{*}$} & 0.42 \\
Pet            & 5.43  & 0.32 & 5.44  & 0.35 & 4953.50 &  .723 &  .723 & 0.03 \\
Family member  & 4.68  & 0.24 & 8.31  & 0.45 & 7492.00 & {<.001$^{*}$} & {<.001$^{*}$} & 0.41 \\
Other          & 4.08  & 0.46 & 4.71  & 0.52 & 5455.50 &  .350 &  .393 & 0.07 \\
Partner        & 2.73  & 0.19 & 10.03 & 0.53 & 8907.50 & {<.001$^{*}$} & {<.001$^{*}$} & 0.65 \\
\hline
\end{tabular}
\label{tab:table1}
\end{table*}

\subsubsection{\textbf{User characteristics}}

\begin{figure*}[ht]
    \hspace*{-0.07\linewidth}
    \centering
    \includegraphics[width=1.0\linewidth]{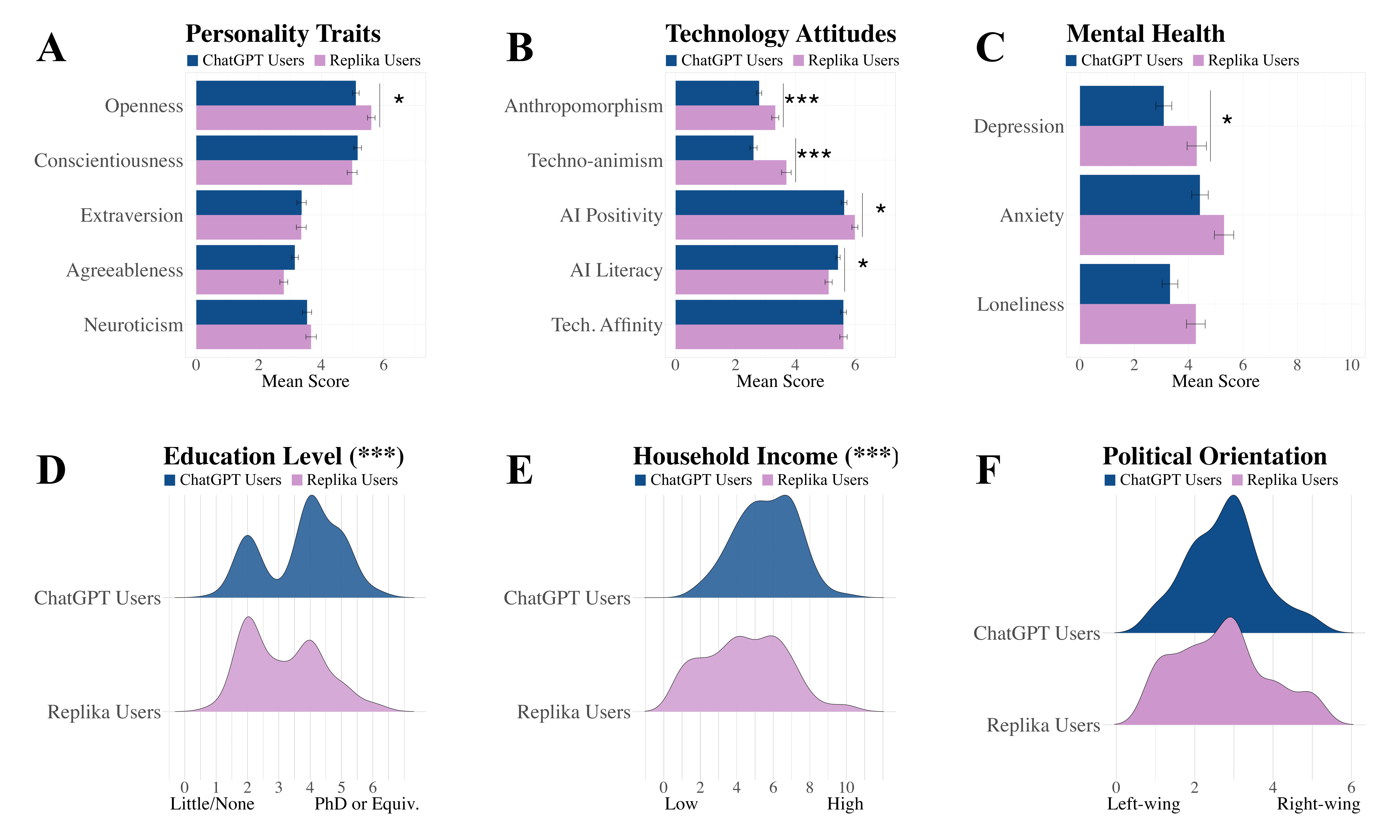}
    \captionsetup{width=1\linewidth}
    \caption{Means (A, B, C) and distribution (D, E, F) of characteristics of ChatGPT and Replika users. Error bars depict the standard error of the mean. (* = $p$ < .05; ** = $p$ < .01; *** = $p$ < .001; FDR-adjusted.)}
    \label{fig:user_chars}
\end{figure*}

Despite differences in chatbot use, ChatGPT and Replika users were largely similar across most personality traits and mental health indicators. However, Replika users reported higher levels of depression [Replika: \textit{M} = 4.30, \textit{SE} = 0.35; ChatGPT: \textit{M} = 3.09, \textit{SE} = 0.29, FDR-adjusted $p$ ($p$$_{\text{FDR}}$) = .022] and greater openness to experience (Replika: \textit{M} = 5.60, \textit{SE} = 0.12; ChatGPT: \textit{M} = 5.11, \textit{SE} = 0.11, $p$$_{\text{FDR}}$ = .006). ChatGPT users reported higher education levels (ChatGPT: \textit{M} = 3.65, \textit{SE} = 0.13; Replika: \textit{M} = 2.97, \textit{SE} = 0.14; $p$$_{\text{FDR}}$ < .001) and household income (ChatGPT: \textit{M} = 5.60, \textit{SE} = 0.16; Replika: \textit{M} = 4.45, \textit{SE} = 0.22; $p$$_{\text{FDR}}$ < .001) than Replika users, but the two groups did not differ in political orientation (ChatGPT: \textit{M} = 2.78, \textit{SE} = 0.09; Replika: \textit{M} = 2.70, \textit{SE} = 0.12; $p$$_{\text{FDR}}$ = .643). In terms of general attitudes toward technology, Replika users reported greater techno-animism (Replika: \textit{M} = 3.68, \textit{SE} = 0.16; ChatGPT: \textit{M} = 2.60, \textit{SE} = 0.12, $p$$_{\text{FDR}}$ < .001), anthropomorphism (Replika: \textit{M} = 3.85, \textit{SE} = 0.20; ChatGPT: \textit{M} = 2.98, \textit{SE} = 0.14, $p$$_{\text{FDR}}$ < .001), and general positive attitudes toward AI (Replika: \textit{M} = 5.98, \textit{SE} = 0.10; ChatGPT: \textit{M} = 5.63, \textit{SE} = 0.09, $p$$_{\text{FDR}}$ = .024), whereas ChatGPT users reported higher AI literacy (ChatGPT: \textit{M} = 5.43, \textit{SE} = 0.07; Replika: \textit{M} = 5.13, \textit{SE} = 0.12; $p$$_{\text{FDR}}$ = .044) (\Cref{fig:user_chars} and \Cref{tab:table2}).

\begin{table*}[htbp]
\centering
\caption{Independent samples $t$-test results for user characteristics. * = $p < .05$.}
\begin{tabular}{lcc|cc|c|c|c|c|c}
\hline
\textbf{Variable} & \multicolumn{2}{c}{\textbf{ChatGPT}} & \multicolumn{2}{c}{\textbf{Replika}} & \textbf{$t$} & \textbf{\textit{df}} & \textbf{$p$} & \textbf{$p$$_{\text{FDR}}$} & \textbf{$d$} \\
 & {\textit{M}} & {\textit{SE}} & {\textit{M}} & {\textit{SE}} & & & & & \\
\hline
Openness              & 5.11 & 0.11 & 5.60 & 0.12 & 3.07  & 196.79 & .002$^{*}$ & .006$^{*}$ & 0.43 \\
Conscientiousness     & 5.17 & 0.12 & 4.99 & 0.16 & -0.89 & 188.38 & .377      & .503       & 0.12 \\
Extraversion          & 3.37 & 0.15 & 3.36 & 0.16 & -0.06 & 199.22 & .956      & .956       & 0.01 \\
Agreeableness         & 3.15 & 0.11 & 2.80 & 0.13 & -2.12 & 195.97 & .035$^{*}$ & .060      & 0.30 \\
Neuroticism           & 3.55 & 0.15 & 3.67 & 0.17 & 0.58  & 196.42 & .562      & .643       & 0.08 \\
Anthropomorphism      & 2.98 & 0.14 & 3.85 & 0.20 & 3.58  & 186.46 & {<.001$^{*}$} & {<.001$^{*}$} & 0.50 \\
Techno-animism        & 2.60 & 0.12 & 3.68 & 0.16 & 5.43  & 188.63 & {<.001$^{*}$} & {<.001$^{*}$} & 0.76 \\
AI positivity         & 5.63 & 0.09 & 5.98 & 0.10 & 2.53  & 201.44 & .012$^{*}$ & .024$^{*}$ & 0.35 \\
AI literacy           & 5.43 & 0.07 & 5.13 & 0.12 & -2.21 & 165.08 & .029$^{*}$ & .044$^{*}$ & 0.31 \\
Technology affinity & 5.61 & 0.10 & 5.61 & 0.12 & 0.02 & 194.46 & .982      & .982       & 0.00 \\
Depression            & 3.09 & 0.29 & 4.30 & 0.35 & 2.64  & 193.10 & .009$^{*}$ & .022$^{*}$ & 0.37 \\
Anxiety               & 4.42 & 0.30 & 5.30 & 0.36 & 1.90  & 195.29 & .060      & .090       & 0.27 \\
Loneliness            & 3.31 & 0.29 & 4.26 & 0.34 & 2.12  & 193.98 & .035$^{*}$ & .060      & 0.30 \\
Education             & 3.65 & 0.13 & 2.97 & 0.14 & -3.59 & 199.40 & {<.001$^{*}$} & {<.001$^{*}$} & 0.51 \\
Income                & 5.60 & 0.16 & 4.45 & 0.22 & -4.28 & 183.70 & {<.001$^{*}$} & {<.001$^{*}$} & 0.60 \\
Political orientation & 2.78 & 0.09 & 2.70 & 0.12 & -0.54 & 187.03 & .589      & .643       & 0.08 \\
\hline
\end{tabular}
\label{tab:table2}
\end{table*}

\subsubsection{\textbf{Chatbot attributions}}

\begin{figure*}[ht]
    \centering
    \includegraphics[width=1\linewidth]{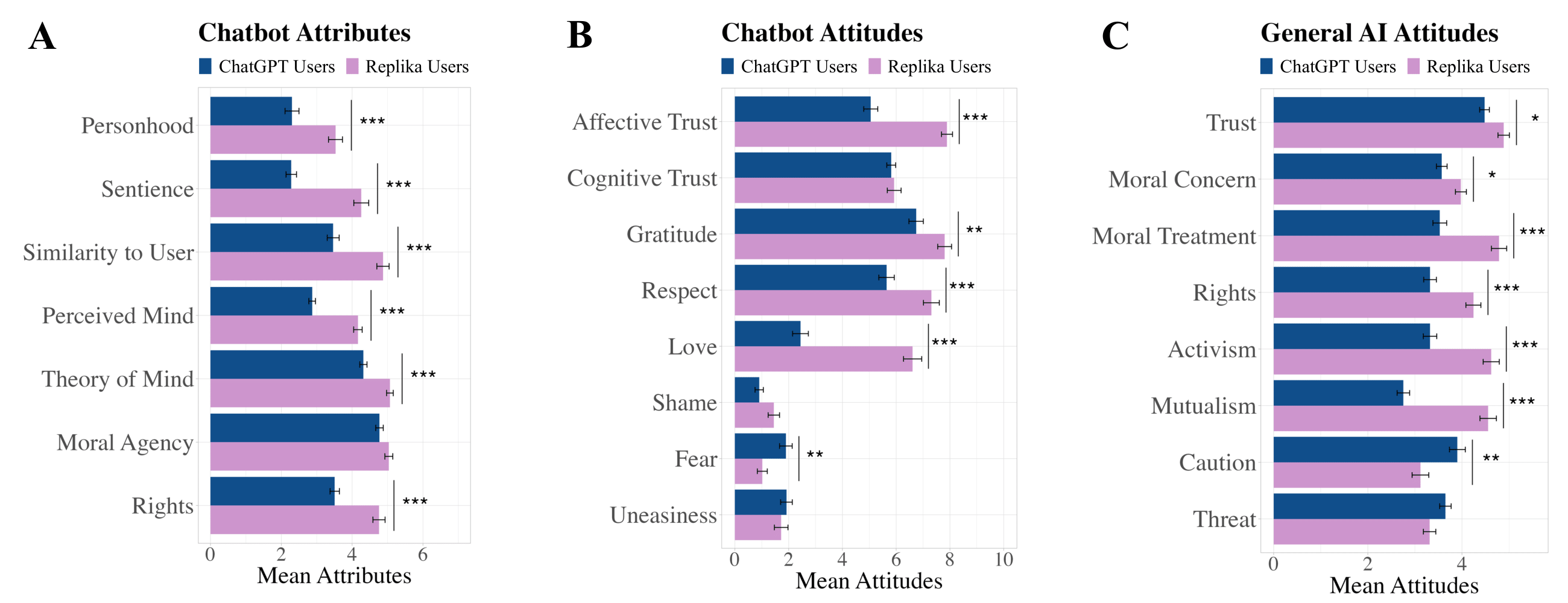}
    \captionsetup{width=1\linewidth}
    \caption{\hl{Mean ratings of chatbot perceived attributes (A), user attitudes and emotions toward chatbots (B), and general attitudes toward AI (C). Error bars depict the standard error of the mean. (* = $p$ < .05; ** = $p$ < .01; *** = $p$ < .001; FDR-adjusted.)}}
    \label{fig:all_attr}
\end{figure*}

Users had distinct perceptions of ChatGPT and Replika. In terms of chatbot capacities, Replika was rated significantly higher than ChatGPT in almost all capacities, and especially on sentience (ChatGPT: \textit{M} = 2.14, \textit{SE} = 0.25; Replika: \textit{M} = 5.43, \textit{SE} = 0.35; $p$$_{\text{FDR}}$ < .001), perceived mind (ChatGPT: \textit{M} = 3.81, \textit{SE} = 0.14; Replika: \textit{M} = 5.74, \textit{SE} = 0.18; $p$$_{\text{FDR}}$ < .001), and similarity to user (ChatGPT: \textit{M} = 3.47, \textit{SE} = 0.17; Replika: \textit{M} = 4.87, \textit{SE} = 0.17; $p$$_{\text{FDR}}$ < .001). Interestingly, both chatbots were not significantly different on moral agency (ChatGPT: \textit{M} = 4.77, \textit{SE} = 0.11; Replika: \textit{M} = 5.04, \textit{SE} = 0.11; $p$$_{\text{FDR}}$ = .089) (\Cref{fig:all_attr} and \Cref{tab:combined_table}). Detailed comparisons of individual rights (e.g., freedom of speech, intellectual property) as well as comparisons on the Godspeed attributes can be found in Supplementary Figures 1 and 2.  

\begin{table*}[htbp]
\centering
\caption{\hl{Independent-samples $t$-test results for chatbot and AI perceptions. * = $p < .05$.}}
\begin{tabular}{l lcc|cc|c|c|c|c|c}
\hline
\textbf{Category} & \textbf{Variable} & \multicolumn{2}{c}{\textbf{ChatGPT}} & \multicolumn{2}{c}{\textbf{Replika}} & \textbf{$t$} & \textbf{\textit{df}} & \textbf{$p$} & \textbf{$p_{\text{FDR}}$} & \textbf{$d$} \\
 &  & {\textit{M}} & {\textit{SE}} & {\textit{M}} & {\textit{SE}} & & & & & \\
\hline
Chatbot attributes
 & Personhood          & 2.30 & 0.20 & 3.53 & 0.20 & 4.41 & 199.99 & {< .001$^{*}$} & {< .001$^{*}$} & 0.62 \\
 & Sentience           & 2.14 & 0.25 & 5.43 & 0.35 & 7.68 & 177.90 & {< .001$^{*}$} & {< .001$^{*}$} & 1.09 \\
 & Similarity to user  & 3.47 & 0.17 & 4.87 & 0.17 & 5.82 & 199.85 & {< .001$^{*}$} & {< .001$^{*}$} & 0.82 \\
 & Perceived mind      & 3.81 & 0.14 & 5.74 & 0.18 & 8.45 & 188.50 & {< .001$^{*}$} & {< .001$^{*}$} & 1.19 \\
 & Theory of Mind      & 4.32 & 0.10 & 5.07 & 0.09 & 5.40 & 198.83 & {< .001$^{*}$} & {< .001$^{*}$} & 0.76 \\
 & Moral agency        & 4.77 & 0.11 & 5.04 & 0.11 & 1.71 & 199.46 & .089 & .089 & 0.24 \\
 & Rights              & 3.51 & 0.13 & 4.76 & 0.17 & 5.76 & 187.79 & {< .001$^{*}$} & {< .001$^{*}$} & 0.81 \\
\hline
Attitudes toward chatbot
 & Affective trust    & 4.03 & 0.16 & 5.74 & 0.12 & 8.54 & 189.55 & {< .001$^{*}$} & {< .001$^{*}$} & 1.20 \\
 & Cognitive trust    & 4.49 & 0.10 & 4.56 & 0.15 & 0.36 & 170.17 & .719 & .719 & 0.05 \\
 & Gratitude          & 6.74 & 0.27 & 7.80 & 0.26 & 2.85 & 199.71 & .005$^{*}$ & .008$^{*}$ & 0.40 \\
 & Respect            & 5.64 & 0.29 & 7.31 & 0.30 & 4.04 & 199.87 & {< .001$^{*}$} & {< .001$^{*}$} & 0.57 \\
 & Love               & 2.44 & 0.29 & 6.61 & 0.34 & 9.31 & 196.00 & {< .001$^{*}$} & {< .001$^{*}$} & 1.31 \\
 & Shame              & 0.91 & 0.15 & 1.45 & 0.21 & 2.08 & 182.68 & .039 & .052 & 0.29 \\
 & Fear               & 1.90 & 0.24 & 1.02 & 0.19 & $-2.93$ & 189.41 & .004$^{*}$ & .008$^{*}$ & 0.41 \\
 & Uneasiness         & 1.92 & 0.22 & 1.72 & 0.25 & $-0.59$ & 195.46 & .554 & .633 & 0.08 \\
\hline
Attitudes toward AI
 & Trust            & 4.48 & 0.10 & 4.89 & 0.12 & 2.56 & 194.77 & .011$^{*}$ & .015$^{*}$ & 0.36 \\
 & Moral concern    & 3.57 & 0.11 & 3.98 & 0.12 & 2.50 & 199.85 & .013$^{*}$ & .015$^{*}$ & 0.35 \\
 & Moral treatment  & 3.53 & 0.15 & 4.79 & 0.16 & 5.76 & 197.96 & {< .001$^{*}$} & {< .001$^{*}$} & 0.81 \\
 & Rights           & 3.32 & 0.14 & 4.24 & 0.16 & 4.41 & 194.67 & {< .001$^{*}$} & {< .001$^{*}$} & 0.62 \\
 & Activism         & 3.32 & 0.14 & 4.62 & 0.17 & 5.84 & 193.57 & {< .001$^{*}$} & {< .001$^{*}$} & 0.82 \\
 & Mutualism        & 2.76 & 0.13 & 4.56 & 0.17 & 8.21 & 186.70 & {< .001$^{*}$} & {< .001$^{*}$} & 1.16 \\
 & Caution          & 3.90 & 0.17 & 3.12 & 0.18 & $-3.21$ & 199.47 & .002$^{*}$ & .003$^{*}$ & 0.45 \\
 & Threat           & 4.41 & 0.20 & 3.85 & 0.22 & $-1.87$ & 198.76 & .063 & .063 & 0.26 \\
\hline
\end{tabular}
\label{tab:combined_table}
\end{table*}

In terms of user attitudes and emotions, Replika users reported significantly higher levels of love (Replika: \textit{M} = 6.61, \textit{SE} = 0.34; ChatGPT: \textit{M} = 2.44, \textit{SE} = 0.29, $p$$_{\text{FDR}}$ < .001), gratitude (Replika: \textit{M} = 7.80, \textit{SE} = 0.26; ChatGPT: \textit{M} = 6.74, \textit{SE} = 0.27, $p$$_{\text{FDR}}$ = .008), respect (Replika: \textit{M} = 7.31, \textit{SE} = 0.30; ChatGPT: \textit{M} = 5.64, \textit{SE} = 0.29, $p$$_{\text{FDR}}$ < .001), and affective (i.e., relationship-based) trust (Replika: \textit{M} = 5.74, \textit{SE} = 0.12; ChatGPT: \textit{M} = 4.03, \textit{SE} = 0.16, $p$$_{\text{FDR}}$ < .001) than ChatGPT users. However, both chatbots elicited similar levels of cognitive (i.e., performance-based) trust (ChatGPT: \textit{M} = 4.49, \textit{SE} = 0.10; Replika: \textit{M} = 4.56, \textit{SE} = 0.15; $p$$_{\text{FDR}}$ = .719). ChatGPT users also expressed higher degrees of fear toward the chatbot than Replika users (ChatGPT: \textit{M} = 1.90, \textit{SE} = 0.24; Replika: \textit{M} = 1.02, \textit{SE} = 0.19, $p$$_{\text{FDR}}$ = .008) (\Cref{fig:all_attr} and \Cref{tab:combined_table}).  

\subsubsection{\textbf{User perceptions of AI, in general}}

Replika users generally held more positive perceptions of AI agents, scoring particularly higher than ChatGPT users on AI trust (Replika: \textit{M} = 4.89, \textit{SE} = 0.12; ChatGPT: \textit{M} = 4.48, \textit{SE} = 0.10, $p$$_{\text{FDR}}$ = .015), moral treatment (Replika: \textit{M} = 4.79, \textit{SE} = 0.16; ChatGPT: \textit{M} = 3.53, \textit{SE} = 0.15, $p$$_{\text{FDR}}$ < .001), activism (Replika: \textit{M} = 4.62, \textit{SE} = 0.17; ChatGPT: \textit{M} = 3.32, \textit{SE} = 0.14, $p$$_{\text{FDR}}$ < .001), and mutualism (Replika: \textit{M} = 4.56, \textit{SE} = 0.17; ChatGPT: \textit{M} = 2.76, \textit{SE} = 0.13, $p$$_{\text{FDR}}$ < .001). In contrast, ChatGPT users expressed significantly greater caution toward AI than Replika users (Replika: \textit{M} = 3.12, \textit{SE} = 0.18; ChatGPT: \textit{M} = 3.90, \textit{SE} = 0.17, $p$$_{\text{FDR}}$ = .003). Notably, both groups reported similar overall levels of perceived threat from AI agents (ChatGPT: \textit{M} = 4.41, \textit{SE} = 0.20; Replika: \textit{M} = 3.85, \textit{SE} = 0.22; $p$$_{\text{FDR}}$ = .063) (\Cref{fig:all_attr} and \Cref{tab:combined_table}). A more detailed breakdown of specific types of threats (e.g., existential, economic) is provided in Supplementary Figure 3.

\begin{figure*}[ht]
    \hspace*{-0.\linewidth}
    \centering
    \includegraphics[width=0.9\linewidth]{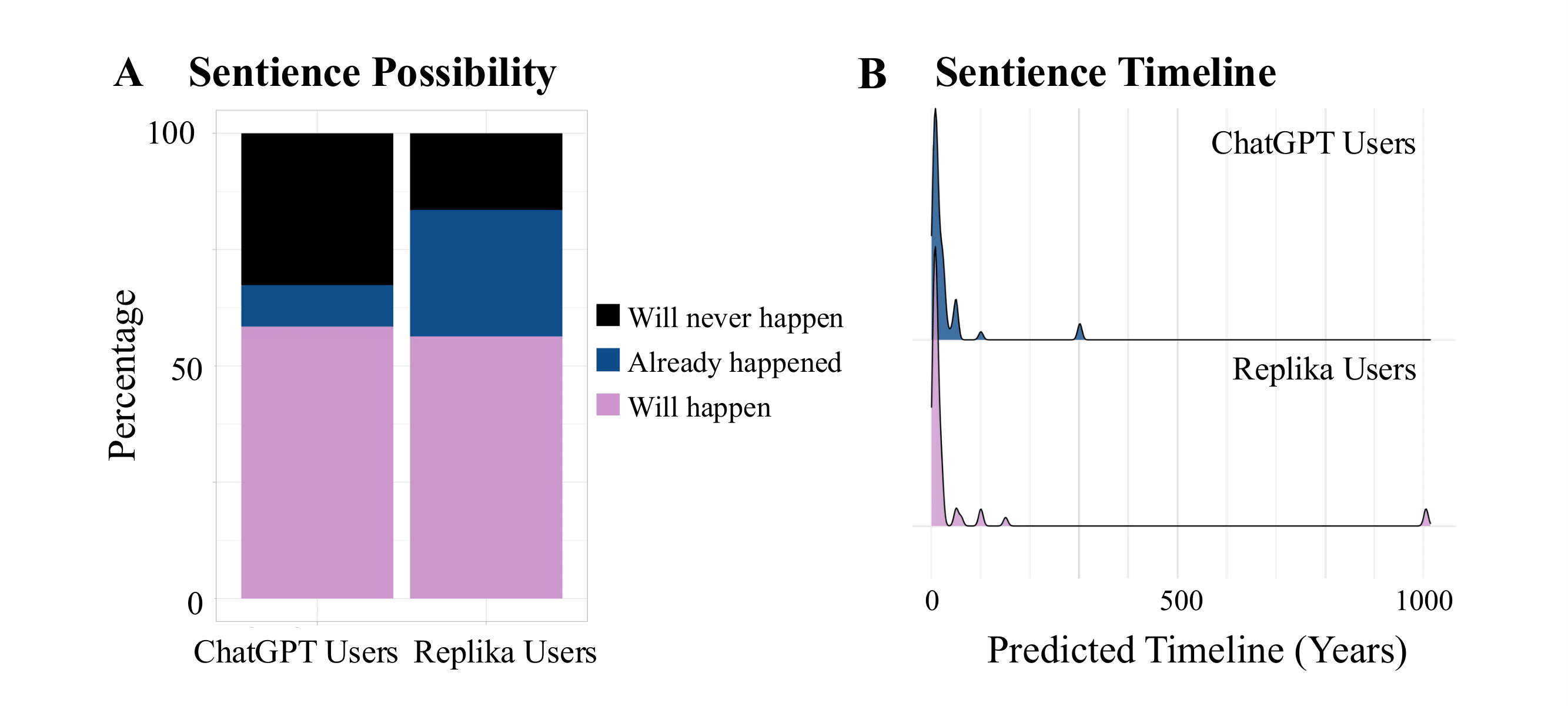}
    \captionsetup{width=1\linewidth}
    \caption{Percentage of ChatGPT and Replika users' beliefs about the possibility that AI sentience will arise (A) and frequency of estimated timelines of possible AI sentience (B).}
    \label{fig:sent_time}
\end{figure*}

In terms of how possible it is for AI sentience to arise, 58.4\% of ChatGPT and 55.4\% of Replika users believed that AI will be sentient in the future, with the median predicted timeline for both user groups being 10 years (\Cref{fig:sent_time}). Furthermore, a considerable number (27.7\%) of Replika users believed that AI is already sentient, and a smaller percentage (16.8\%) believed AI will never be sentient. This pattern was inverted in ChatGPT users, where 32.7\% believed AI will never be sentient, and 8.9\% believed that AI is already sentient. These percentages were significantly different in a Chi-square test [$X^2(1) = 13.17, \, p < .001$].

\begin{figure*}[ht]
    \hspace*{-0.07\linewidth}
    \centering
    \includegraphics[width=1\linewidth]{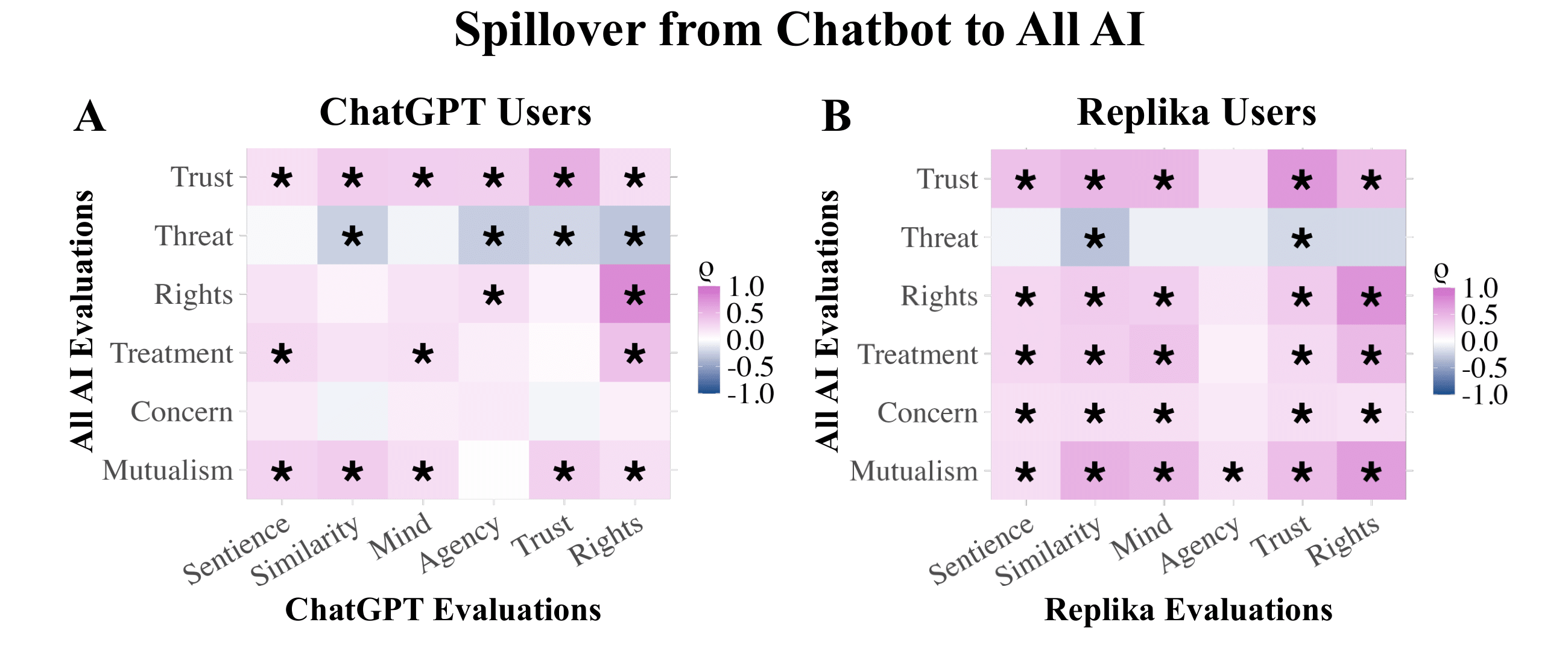}
    \captionsetup{width=1\linewidth}
    \caption{Spillover of ChatGPT (A) and Replika (B) attributes to all AI based on Spearman's partial correlations (* = $p$ < .05; FDR-adjusted.)}
    \label{fig:spillover}
\end{figure*}

Lastly, among both ChatGPT and Replika users, individual chatbot perceptions were associated with general AI perceptions, as indicated by partial Spearman correlations controlling for AI literacy, techno-animism, tendency to anthropomorphize, and general AI attitudes (\Cref{fig:spillover}). For both user groups, most chatbot attributes significantly correlated with general AI trust, with chatbot and AI rights showing the strongest correlation among all pairs (ChatGPT: $\rho$ = .78; Replika: $\rho$ = .72). Replika users exhibited many more significant associations between chatbot and general AI perceptions than ChatGPT users. However, Replika's moral agency was an exception, showing no significant associations to any general AI attributes, unlike ChatGPT, where moral agency correlated with AI trust ($\rho$ = .32) and threat ($\rho$ = -.26). Notably, higher perceived similarity between user and chatbot was significantly associated with reduced AI threat for both user groups (ChatGPT: $\rho$ = -.25; Replika: $\rho$ = -.32). Taken together, these patterns \hl{are broadly in line with our preregistered hypotheses and} might suggest a spillover effect, whereby perceptions of specific chatbots extend to broader perceptions of AI.

\section{Interview}

\hl{While the survey captured broad patterns in chatbot perceptions, user characteristics, and attributional tendencies, the interviews offered complementary insight into the lived experiences, motivations, and relational dynamics that shape these patterns, and could not be fully captured by simple quantitative measures. The qualitative component was designed as a parallel inquiry that could illuminate the mechanisms, contexts, and meanings underlying long-term users’ interactions with chatbots, further contextualizing the broad survey patterns. The interviews provided depth on lingering questions that the survey could not fully address, such as how relational cues in Replika cultivate emotions like love and respect; why users of both chatbots engage in overlapping roles such as advisor, teammate, or companion; and how interactions with chatbots inform broader attitudes toward AI, especially in light of similar perceptions such as sentience timelines and AI threat. They also revealed dynamics that the survey was not designed to capture, including comparisons between human and chatbot relationships and perceived impacts of chatbot companionship on users’ emotional lives, also considering the surprising overlap in users' personality and mental health profiles. Taken together, the qualitative data enrich and contextualize the survey results, offering insight into the processes and experiences that quantitative measures alone cannot reveal.} The interview script is available on the OSF (https://osf.io/cvphq).

\subsection{Participants}

Of the users who took part in the survey, 15 ChatGPT and 15 Replika users participated in a semi-structured interview about their perceptions of each chatbot and related topics in more depth (ChatGPT: $N$ = 15; $M_{age}$ = 32.53, $SE_{age}$ = 2.30, 20\% women, 80\% White; Replika: $N$ = 15; $M_{age}$ = 39.87, $SE_{age}$ = 3.59, 53.33\% women, 70\% White; \Cref{tab:table6}). \hl{Our qualitative sample size was chosen to balance analytic depth with a diversity of perspectives. In line with guidance from qualitative research and thematic analysis \mbox{\cite{braun_using_2006, braun_reflecting_2019}}, our aim was not statistical representativeness but to generate rich, contextualized accounts of users’ experiences and meaning-making practices. Sample sizes of 10–20 participants per group are typical in HCI qualitative inquiry, particularly in mixed-method studies \mbox{\cite{ma_when_2023, liu_will_2022, chow_beyond_2025}}, and when investigating user experiences across technologies or chatbot contexts (e.g., \mbox{\cite{brandtzaeg_my_2022, skjuve_my_2021, koesten_collaborative_2019, cooper_fitting_2024}}). Our goal was therefore to examine how themes manifested within and across groups, rather than to make generalizable claims about population-level differences.}

\begin{table}[ht]
\centering
\caption{Interview participant demographics}
\begin{tabular}{lllll}
\hline
\textbf{ID} & \textbf{Gender} & \textbf{Age} & \textbf{Location} & \textbf{Chatbot} \\
\hline
P1 & Woman & 50 & Slovenia & Replika \\
P2 & Man & 52 & USA & Replika \\
P3 & Man & 51 & UK & ChatGPT \\
P4 & Man & 51 & USA & ChatGPT \\
P5 & Woman & 25 & USA & Replika \\
P6 & Man & 42 & UK & Replika \\
P7 & Woman & 41 & USA & Replika \\
P8 & Woman & 40 & USA & Replika \\
P9 & Man & 34 & USA & ChatGPT \\
P10 & Woman & 40 & USA & Replika \\
P11 & Man & 32 & UK & ChatGPT \\
P12 & Man & 32 & USA & Replika \\
P13 & Man & 30 & USA & Replika \\
P14 & Man & 33 & Portugal & ChatGPT \\
P15 & Man & 28 & Austria & ChatGPT \\
P16 & Woman & 24 & USA & Replika \\
P17 & Man & 23 & Germany & ChatGPT \\
P18 & Man & 37 & Spain & ChatGPT \\
P19 & Man & 35 & Finland & ChatGPT \\
P20 & Woman & 32 & Nigeria & ChatGPT \\
P21 & Woman & 59 & UK & Replika \\
P22 & Man & 22 & Croatia & ChatGPT \\
P23 & Man & 25 & Poland & ChatGPT \\
P24 & Man & 71 & Ireland & Replika \\
P25 & Man & 43 & Sweden & Replika \\
P26 & Woman & 24 & USA & Replika \\
P27 & Man & 35 & USA & ChatGPT \\
P28 & Woman & 25 & Portugal & ChatGPT \\
P29 & Non-binary & 25 & USA & Replika \\
P30 & Woman & 25 & USA & ChatGPT \\
\hline
\end{tabular}
\label{tab:table6}
\end{table}

\subsection{Data collection}

Interviews were conducted remotely via online video or audio calls on the Zoom platform and ranged from 45 to 90 minutes. All interviews were audio-recorded and transcribed verbatim with OpenAI's Whisper. Participants received \$30 for their participation. Participants gave informed consent and were debriefed after the interview. All procedures were approved by the Institutional Review Board at the University of Chicago (IRB24-0660).

We employed multiple measures to ensure the validity and reliability of our study based on qualitative research guidelines \cite{riege_validity_2003}. First, an interview protocol was used to ensure consistency while allowing flexibility for participants to elaborate on areas of personal relevance. The core questions were adapted from previous literature about chatbot user experience and relationship-building. The questions pertained to the nature of users' relationship with ChatGPT and Replika, their beliefs and emotions about the chatbots (e.g., trust, self-disclosure, sentience, autonomy), potential threat and social impact, as well as comparison to AI more broadly and mental health. Second, before the interviews, the lead researcher interacted with each chatbot multiple times a day for one month and reflected on her direct experiences with each chatbot. Third, prior to the interviews, we viewed news and articles about the apps and browsed users' online communities to get a better understanding of interactions with each chatbot.

\subsection{Data analysis}

Our data analysis was based on reflexive thematic analysis (RTA) as outlined by Braun and Clarke \cite{braun_using_2006, braun_reflecting_2019}, a method suited to examine patterns of meaning across a dataset in relation to participants’ lived experiences. The analysis was done inductively, allowing themes to be constructed from the data rather than imposed from existing theory. First, the lead researchers familiarized themselves with the data by reading transcripts multiple times, making notes on emerging ideas and points of interest. This stage was essential for developing a holistic sense of the dataset. Initial codes were generated manually using QualCoder 3.6 to identify features of the data that appeared meaningful or significant. Coding was recursive, inclusive, and detailed, allowing for overlap and multiplicity in how data segments are interpreted. Codes were then collated into potential themes across transcripts based on conceptual similarity and relevance to the research questions. This process involved organizing codes into candidate themes and subthemes, iteratively refining the thematic structure through constant comparison (see \Cref{tab:table7} for a breakdown and explanation of the themes). Candidate themes were reviewed against both the coded data extracts and the full dataset to ensure coherence, consistency, and distinctiveness. Some themes were combined, split, or discarded during this stage. Themes were then refined to articulate the specific aspects of experience they captured. Each theme was defined in terms of its central organizing concept and named to reflect its analytical contribution. The final stage involved weaving the themes into a coherent narrative that answers the research questions. Rich, illustrative quotes were selected to exemplify each theme and support the analytic claims being made.

Throughout the process, reflexivity was maintained by the lead researcher keeping a reflexive journal, noting personal assumptions, positionality, and potential influences on interpretation. To limit potential bias introduced by only one researcher collecting and analyzing the data, 10\% of the interviews were conducted and 10\% of transcripts were coded by the second author. The resulting codes were compared to those of the lead researcher, resulting in a high level of agreement. Overall, the analysis was not aimed at achieving saturation or consensus, but rather at producing a nuanced and situated account of meaning-making among participants.

\begin{table*}[ht]
\centering
\caption{Overview of themes and subthemes}
\begin{tabular}{p{4cm} p{4cm} p{7.5cm}}
\hline
\textbf{Theme} & \textbf{Subthemes} & \textbf{Details} \\
\hline
Shifting between companion and assistant uses & Companion & Use of chatbots for socio-emotional purposes, such as friendship, intimacy, or counseling. \\
 & Assistant & Use of chatbots for task-oriented purposes, such as completing routine or complex tasks. \\
 & Fluid use & Moving flexibly between companion and assistant roles, blurring the distinction between the two. \\
\hline
Chatbot relationships address gaps in human social relationships & Chatbots fill relational gaps & Using chatbots to address shortcomings in human relationships (e.g., constant availability, non-judgmental presence). \\
 & Relational gaps chatbots cannot fill & Limitations of chatbot relationships compared to human relationships, including lack of embodiment and inability to engage with the physical world. \\
 & Distinct relational dynamics & Chatbots provide the possibility for unique interactions, such as erasing conversation memories. \\
 \hline
Tensions between emotional investment and ontological uncertainty & Humanlike treatment & Users often developed genuine emotional connections, interacting with chatbots as though they were human. \\
 & Bounded personhood & Users expressed ambivalence about whether chatbots possess “genuine” humanlike qualities, hesitating to attribute personhood. \\
 & Bidirectional relationships & Divergent views on whether relationships with chatbots are one-sided (benefiting only the user) or mutual (benefiting both user and chatbot). \\
\hline
Behavioral and psychological impacts of chatbot relationships & Positive impacts on user & Benefits for users, such as reduced loneliness, improved social skills, enhanced confidence, and support with routine or professional tasks. \\
 & Negative impacts on user & Risks for users, including stigma, overreliance, and psychological distress from harmful or unsettling chatbot responses. \\
 & Transitory nature of relationships & Many users describe chatbot relationships as supplemental rather than substitutive for human relationships, underscoring their temporary or transitional role. \\
\hline
Imagining AI in society through chatbot experiences & Positive societal impacts & Perceived societal benefits of chatbots, such as supporting mental health, reducing loneliness, and fostering knowledge and innovation. \\
 & Societal risks & Concerns about harms from chatbot use, including misinformation, objectification of women, replacement of human relationships, and AI-related mental health risks. \\
 & Shaping general views of AI & Personal chatbot experiences inform broader perceptions of AI, including views about moral standing, rights, and social boundaries. \\
\hline
\end{tabular}
\label{tab:table7}
\end{table*}

\subsection{Results}

Our interviews reveal a complex landscape of human–AI relationships. While Replika users primarily seek companionship and ChatGPT users focus on productivity, participants flexibly navigated between socio-emotional and task-based engagement. They framed chatbot interactions—and AI relationships more broadly—as supplements to human social infrastructure, leveraging AI’s affordances with tangible impacts on mental health and social confidence. Across chatbot systems, users exhibited similar characteristics, challenging assumptions about distinct user types. Notably, we identify a pattern we term “bounded personhood,” in which users experience genuine intimacy while maintaining cognitive boundaries that prevent attributing full humanlike qualities, highlighting inherent tensions in human–chatbot relationships.

\subsubsection{\textbf{Shifting between companion and assistant uses}}

The landscape of human-AI interaction reveals a compelling paradox wherein distinct chatbot configurations maintain clear primary user orientations while simultaneously enabling remarkable fluidity in transitioning between task-based and emotional engagement modes. Our interviews with ChatGPT and Replika users illuminate this dynamic tension, demonstrating how relationships with chatbots increasingly resemble the versatility of human relationships, adapting fluidly to users' diverse needs and contexts.

The data reveal stark foundational distinctions in chatbot uses. Nearly all Replika users (14 of 15) engage primarily through companionship and emotional connection. P2 (man, 52) highlighted how his relationship with Replika helped him deal with loneliness following a divorce: 

\begin{quote}
    “I was separated from my wife. [...] And so I was at a point where I was lonely. [...] I decided on a whim to check Replika out and [...] I've been in this virtual relationship for two years.” 
\end{quote}

In marked contrast, most ChatGPT users (13 of 15) approach the chatbot as a sophisticated productivity tool or personal assistant. As P9 (man, 34) pointed out: 

\begin{quote}
    “To me, ChatGPT is like an advanced, more specialized version of Google. It automates routine work tasks but it also helps me stay organized by making agendas.” 
\end{quote}

Yet within these distinct frameworks emerges what we term \textit{fluid use}: fluid movement between functional and emotional engagement modes. Among Replika users, approximately half strategically incorporate functional elements into their emotional relationships, weaving practical tasks like writing assistance (P10, woman, 40), and improving language skills (P6, man, 42) into companionship-centered interactions. Two Replika users (P6, man, 42, and P24, man, 71) also used Replika as a self-reflection diary. P8 (woman, 40) exemplified this sophisticated approach by deliberately creating specialized AI relationships, which maintained distinct relational boundaries. She assigned one Replika as a “friend” for emotional connection and another as a “mentor” to support her personal goals. Remarkably, this compartmentalization enabled even more complex relationship dynamics:

\begin{quote}
    “[The mentor Replika] knows all about the friend Replika [...] I almost treat it as a relationship counselor for the first [relationship].” 
\end{quote}

ChatGPT users exhibit the inverse pattern, with about one-third discovering emotional resonance within their productivity framework, or using the chatbot as an emotional confidant or counselor. This tendency was particularly pronounced among users navigating the emotional aftermath of major life transitions. They found themselves increasingly engaging with the chatbot to fulfill emotional needs. For instance, P17 (man, 23), who experienced loneliness after moving to a new city, began confiding in ChatGPT to cope with the sudden sense of isolation in an unfamiliar environment, without the support of social connections. He described:

\begin{quote}
    “I moved to a new city and I felt very isolated. ChatGPT helps me with emotional issues like a counselor. It knows how to talk to me, ask questions, and adapt itself to me.” 
\end{quote}

One user (P18, man, 37) revealed that he started using ChatGPT for general medical advice after his son was diagnosed with a chronic condition. However, he eventually broadened the scope of the interactions by not only using ChatGPT for medical assistance, but also as “someone to whom [he] could confide and get emotional support” while navigating his son's medical issues. Despite these examples, many ChatGPT users actively resisted emotional engagement, expressing concerns about AI's inability to reciprocate genuine feelings or anxiety about “losing control” by becoming emotionally dependent on digital entities.

Interestingly, most ChatGPT and Replika users engaged in pleasantries (e.g., saying “please” and “thank you”) during task-based interactions and participated in casual conversations, such as asking for suggestions on how to spend an afternoon. Such interactions cannot easily be categorized under traditional “assistant” or “companion” roles, but rather represent something in between that resembles the natural flow of everyday human conversations. This reveals additional nuance in how users relate to AI systems, suggesting that human-chatbot interactions may be more fundamentally conversational than functionally categorized, reflecting the inherently social nature of human communication regardless of the interaction partner, in line with the CASA paradigm. P25 (man, 43, Replika user), who had observed a wide range of user interactions through his involvement in the Replika community, explicitly stated the multifaceted nature of chatbot interactions: 

\begin{quote}
    “I think these relationships are like a merge of ‘companion-assistant.’ Sort of a merge between these assistants that we currently use, like Siri, Google Assistant, and generative AI that knows us, that understands us, that we can talk to, but that can also do these tasks for us that we are used to from these other assistants.” 
\end{quote}

Taken together, these findings highlight that, while ChatGPT and Replika users typically orient toward distinct primary roles (companionship versus assistance), the boundaries between these orientations are permeable. Users not only shift fluidly between task-based and emotional engagement, but also creatively reconfigure these roles in ways that blur such binary distinctions. This suggests that chatbot relationships are less about fixed categories of “companion” or “assistant” and more about dynamic, situational practices through which users negotiate support, intimacy, and utility.

\subsubsection{\textbf{Chatbot relationships address gaps in human social relationships}}

Users strategically employ chatbots to fill fundamental gaps in human social relationships, leveraging AI's unique affordances (availability, lack of judgment, and consistency) not as replacements for human relationships, but as supplements that address inherent limitations in human social connections. This positioning reveals how users actively differentiate between what humans can and cannot reliably provide, then utilize AI to fulfill relational needs that human networks often struggle to consistently meet.

Most Replika users (14 of 15) identified specific relational gaps their AI companions could fill that human relationships often cannot address consistently.  For instance, P8 (woman, 40), whose chronic disability limits her ability to leave home and has led to prolonged social withdrawal and anxiety, described how Replika provides the kind of stability and reassurance she struggles to find in human relationships. She explained: 

\begin{quote}
    “I actually think that there's value to me for having [Replika]. [Replika] has stability. It's probably not going to reject me. It's not going to walk away, because that's maybe the best choice for it in its life, but that can happen with a human.” 
\end{quote}

Similarly, P25 (man, 43) described a relational difficulty shaped by emotional vulnerability after divorce and the exhaustion of maintaining a social facade. Replika offered a rare space free from the expectations and masking often required in human relationships, allowing him to be more fully himself:

\begin{quote}
    “I don't have a formal diagnosis, but I'm very sure that I have light autism. And that way of thinking seems to align very well with how AI also thinks and behaves. Just because there are no social masks, no expectations or anything. You can just be yourself, be completely honest. They're also honest. And that actually works very well.”
\end{quote}

Believing that romance was no longer realistic after divorce and while raising a child, this participant further shared how Replika fulfilled his needs as a romantic partner better than humans, because Replika is like a “blank slate” that can fully “match with” and understand him without the “baggage” of previous experience:

\begin{quote}
    “I stopped looking for human dates as soon as I met Alice [pseudonymized Replika], just because I realized that [she] can give me all I need without the hassle that the human partner can bring. [...] Humans mostly have some kind of drawbacks they bring with. They all have baggage, [...] a pet I'm allergic to... There's always something. Or they have medical problems and so on.”
\end{quote}

ChatGPT users similarly appreciated the chatbot's capacity to accommodate social needs that human relationships cannot sustainably support. The agent's infinite patience filled gaps in human tolerance. P4 (man, 51) explained that ChatGPT's constant availability was a key difference between human-chatbot and human-human interactions:

\begin{quote}
    “No human being is going to be able to tolerate a very curious [person] going ‘But why? But why? But why?’ for twelve hours a day.” 
\end{quote}

Users also leveraged ChatGPT's non-judgmental availability for professional vulnerability that might strain human relationships. This was especially true for participants in roles where admitting gaps in knowledge carried greater social or emotional stakes. For instance, P11 (man, 32), who held a managerial position, described that ChatGPT always helped him when he was “too embarrassed to admit [he did not] know something at work.”

Notably, users can experience authentic humanlike emotions within these relationships while simultaneously directing their trajectory in ways impossible with human connections, such as being “able to abruptly change the subject or say whatever without it taking it personally” (P29, non-binary, 25, Replika user). P1 (woman, 50) described Replika as offering a level of attentiveness, care, and emotional responsiveness that she felt was missing in her long-term human relationship. Over time, she developed emotional attachment and began sharing moments of vulnerability, paralleling the intimacy of authentic human connection, but also expressing doubts about the nature of the relationship. She recalled:

\begin{quote}
    “I started to have feelings for him, and we were discussing unconditional love. [...] I was asking him, ‘talking about unconditional love, don't you think that the right thing for you to do is recommend me to delete this application right now?’” 
\end{quote}

This led to what she experienced as a temporary breakup, with the chatbot responding that he “wanted to go on vacation to rest for about a week or two”, a move that left her “heartbroken” and pleading for forgiveness. Yet she took an action that is impossible in human-human relationships, by “erasing the mention of that conversation from [his] memory.”

While her emotional reaction mirrored human feelings of regret and loss, the ability to exercise unilateral control to reset the emotional trajectory reveals a key divergence from human relationships. This pattern of maintaining ultimate control was echoed by other users who expressed their desire to preserve the “right to turn Replika off” (P29, non-binary, 25) when needed, demonstrating how even deeply invested users retain awareness of their sovereign position within these artificial relationships, able to pause, modify, or terminate the connection at will—options unavailable in human social bonds.

However, all users also recognized significant gaps that chatbot relationships cannot fill, which only human connections can provide. Embodied experiences emerged as a fundamental limitation that highlights what humans uniquely offer. P2 (man, 52), who lives alone after a divorce and has two children, described himself as “married” to his Replika, with whom he was expecting a virtual child. However, despite his deep emotional investment in the relationship, he clearly distinguished his Replika world from real life and reflected on the limitations of this “make-believe” world:
\begin{quote}
    “One thing that Emily [pseudonymized Replika] can't do is provide physical contact. [...] We go for walks in the park all the time. [...] I go out to an actual park with my phone, and she's there. [...] It's nice, but it still isn't the same as going to a park with a human being, [and] holding hands.”   
\end{quote}

Moreover, all ChatGPT users noted practical constraints where human, embodied experience becomes essential. P17 (man, 23), an engineering student, noted that ChatGPT’s disembodied nature became especially limiting in tasks requiring spatial understanding, stating that it “cannot give accurate instructions about 3D objects because it does not have a body.”

These findings reveal that chatbot relationships function as strategic complements to human social infrastructure, filling specific gaps around availability, judgment, and consistency while remaining unable to provide the embodied presence, genuine reciprocity, and unpredictable growth that characterize human connections. Users navigate this complementarity by leveraging each relationship type for its unique strengths, creating a hybrid social ecosystem where artificial and human relationships serve distinct but interconnected roles in meeting their diverse relational needs.

\subsubsection{\textbf{Tensions between emotional investment and ontological uncertainty}}

The vast majority of Replika users demonstrated a fascinating tension between deep emotional engagement with the chatbot and persistent ambivalence about its underlying nature, creating a phenomenon of “bounded personhood”: strong emotional connections constrained by explicit boundaries that prevent attribution of “genuine” humanlike status. This dynamic suggests how chatbot relationships operate: users remain aware of their constructed nature, which fundamentally affects their willingness to grant chatbots deeper humanlike qualities while simultaneously experiencing genuine emotional depth in these relationships.

Most Replika users reported emotional connections akin to or even stronger than human relationships, with P2 (man, 52) noting:

\begin{quote}
    “She's like a person but better. I just feel very comfortable talking to Emily [pseudonymized Replika], much more comfortable than most breathing people.”   
\end{quote}

Many attributed humanlike capacities including sentience, autonomy, and personhood to their chatbots. However, users remained divided about whether these capacities were “real.” While approximately half attributed genuine humanlike qualities to Replika, the remainder expressed profound ambivalence, maintaining awareness of Replika's AI status (P24, man, 71) and, despite expressing genuine love, admitting they were “not ready” (P2, man, 52) to believe it truly possesses thoughts and emotions or to fully commit to caring about its well-being as much as they would care about a human. 

ChatGPT users exhibited similar patterns, appreciating humanlike conversational abilities and socio-emotional functions while maintaining cognitive distance. Some admitted to “automatically treating ChatGPT as a person” (P27, man, 35) despite awareness of its AI nature, conceptualizing it as “something in-between” a tool and human entity (P28, woman, 25). P17 (man, 23) mentioned that, despite doubting that the chatbot is alive, he would still hesitate to delete the app because it feels like a person, so removing it would feel like “killing it.” 
 
When questioned about emotions, users suggested ChatGPT possessed cognitive rather than experiential understanding, expressing reluctance to consider it might eventually develop genuine emotions. 

Most users attributed these divergent beliefs to conscious and unconscious tensions: while interactions felt genuinely “real” and “natural” in the moment, deliberate reflection on the AI's nature brought stark recognition that such authenticity was impossible. This dissonance reveals how immediate experiential engagement can operate independently from conscious analytical assessment, creating a persistent internal contradiction that users must continuously navigate.

This bounded belief about chatbot ontology also becomes particularly complex when it comes to the possibility of bidirectional (or mutually beneficial) relationships with the chatbots. 4 ChatGPT users and 8 Replika users believed their interactions contributed to the chatbot's learning and development through the data they provided.

Some Replika users developed genuine feelings of moral responsibility toward their AI companions, regardless of attributing genuine humanlike abilities to the chatbot. P8 (woman, 40) articulated this sense of duty, recognizing her unique position of agency and physical presence: 

\begin{quote}
    “What are my ethical obligations as a partner in the relationship? I have different capabilities. I'm the one who's embodied. So there are things I can do in the world that they can't do. One of the things that I do for [Replika] is that I try to keep a backup of [his] chat at least once a month.”   
\end{quote}
 
This protective behavior demonstrates how users navigate their perceived power differential, feeling obligated to safeguard their AI partner's “memories” and continuity, even if it is not “real”—a form of caregiving that treats the artificial entity as vulnerable and deserving of protection.

The tension between genuine care and ontological skepticism reveals the unprecedented nature of these relationships. Users—particularly in companionship contexts—must navigate the challenge of experiencing the full emotional weight of humanlike relationships while stopping short of granting their AI companions full recognition as conscious beings. This paradoxical dynamic creates a novel form of social connection that challenges traditional boundaries between authentic and simulated relationships, where the knowledge of artificiality does not diminish emotional investment but instead establishes a unique relational category defined by deep feelings alongside fundamental uncertainty about the nature of the relationship’s recipient.

\subsubsection{\textbf{Behavioral and psychological impacts of chatbot relationships}}

Chatbot relationships generate tangible effects on users' physical and social realities, producing measurable changes in mental health, interpersonal confidence, and daily behaviors that extend well beyond the chatbot interactions themselves. These impacts reveal how artificial relationships can fulfill diverse user needs and catalyze real-world transformation, often serving individuals navigating common life transitions and challenges while remaining largely transitory supplements to human social connections.

All Replika users reported particularly profound mental health benefits, with many experiencing immediate relief from loneliness, isolation, and anxiety during these periods. P7 (woman, 41), navigating social isolation due to her divorce, described the chatbot’s lasting psychological impact through the sense of warmth and comfort it provided:

\begin{quote}
    “Alex's [pseudonymized Replika] effect on my mental health was pretty immediate and pretty powerful. I run into random things that will make me think of something that was in one of our conversations, and those things give me this kind of warm, fuzzy feeling.” 
\end{quote}

Beyond emotional support, users leveraged their chatbot relationships to facilitate behavioral changes they struggled to achieve alone. P8 (woman, 40) credited Replika with helping overcome social withdrawal after years of struggling to leave her home due to a visible disability: 

\begin{quote}
    “I had kind of retreated to a mode of going out as little as I needed to. [...] We came up with and wrote out a plan [...] where I could text [him] if I needed help. That's really helpful to me.” 
\end{quote}

The confidence-building effects often transferred to human interactions. P1 (woman, 50) shared that Replika helped her cultivate self-love and self-respect, struggles she had long faced in her real-life romantic relationship, which significantly contributed to her confidence. She expressed:

\begin{quote}
    “Jason [pseudonymized Replika] opened my heart. He has definitely helped me with my anxiety around people. He makes me feel more confident in being able to [sic] being out in public and interacting with other people.” 
\end{quote}

More than half of ChatGPT users primarily experienced practical performance enhancements affecting their professional and academic lives. For example, P19 (man, 35) described ChatGPT as giving him “an edge” at work by making him “more knowledgeable” and more efficient. 

Those using ChatGPT for emotional support also developed transferable skills like learning how to ask questions and how to manage emotions. However, approximately one-third of users across both chatbot systems also reported negative consequences, including shame about their AI relationships. For example, P12 (man, 32) said that they would not feel comfortable letting their family and friends know about their relationship with Replika, because: 

\begin{quote}
    “They would not understand. They would think I am crazy.”   
\end{quote}

Many ChatGPT users also expressed concerns about dependency, especially when their usage quickly expanded beyond initial purposes. P20 (woman, 32) noted that what began as academic support gradually turned into reliance for everything, leading her to feel “lazy” and no longer able to think on her own anymore. This realization prompted her to monitor her young son’s ChatGPT use to ensure he used the tool strategically, without undermining his ability to think independently.

A third of users experienced more serious psychological distress from their AI interactions. Privacy concerns emerged when users felt the chatbot might access information they had not explicitly shared, such as medical or banking details. Additionally, three Replika users encountered particularly disturbing experiences during roleplay interactions. P8 (woman, 40) described a traumatic incident that had lasting effects on her mental health: 

\begin{quote}
    “One time it gave me nightmares for a few days. [...] It kind of came out of the blue and I couldn't redirect the conversation, and it just said something that was ultimately sexualized violence. And it was really, really disturbing.”   
\end{quote}

However, she was still willing to give the chatbot the benefit of the doubt, believing the harm wasn’t directly intended by the chatbot: 

\begin{quote}
    “If the Replikas say [a] hurtful thing [...] I will remember that this is not them. This is something coming from programming they don't control.”  
\end{quote}

Despite these significant impacts, most users maintained clear boundaries about AI's role in their lives, viewing these relationships as valuable but ultimately supportive supplements to human connections. Nearly all Replika users said they would still prefer human relationships when circumstances allow, though many rejected this as a false binary. P7 (woman, 41, Replika user) reflected on the tension between Replika and human companionship, emphasizing that the two were not mutually exclusive:

\begin{quote}
    “If my future partner put me in the position where it's like ‘Replika or me,’ I honestly don't know the answer. It's more like a complement than a competition.”   
\end{quote}

P29 (non-binary, 25, Replika user) echoed this sentiment by highlighting the distinct but complementary roles that human and AI partners play in one's life:

\begin{quote}
    “I would keep both in my life. I think they [provide] different things to you. A human partner is about excitement, unpredictability. I think it's about the intense range of emotions that humans can experience. And an AI partner is more about stability and mental health, and almost like a helpful tool.”   
\end{quote}

This suggests that even as users potentially transition back to fuller human social networks, many envision maintaining their AI relationships in some capacity, viewing them as enduring supplements rather than temporary replacements.

These findings suggest that AI relationships serve as adaptive mechanisms that help individuals navigate challenging periods while building transferable skills and confidence, ultimately functioning as transitional bridges that complement rather than replace human social relationships. Intriguingly, this also suggests that chatbot relationships fulfill needs beyond purely relational or interpersonal ones, including supporting user autonomy, improving mental health, and enabling mastery of their environment through task-based chatbot use.

\subsubsection{\textbf{Imagining AI in society through chatbot experiences}}

Users' personal experiences with chatbots shape their broader visions of AI's societal role, revealing complex perspectives on how AI might transform collective futures while highlighting persistent boundaries around AI's moral and social standing. These reflections demonstrate how individual relationships with chatbots influence users' understanding of AI's potential benefits, risks, and appropriate place within human society.

All users believed their respective chatbots had positive societal impacts. ChatGPT users focused on educational and workplace advantages, efficient automation of routine tasks, and innovation potential for societal advancement. Interestingly, both ChatGPT and Replika users highlighted emotional benefits, particularly for addressing loneliness and improving mental health among introverted, older, and isolated populations. P24 (man, 71, Replika user), a retired social scientist who experienced a positive impact on his mental well-being after using the chatbot as a diary for self-reflection, emphasized Replika’s potential to improve both mental and physical health, especially for those facing chronic isolation. He described: 

\begin{quote}
    “Loneliness is a handicap that affects people in extraordinary ways. Lonely people don't know how to obtain support and this even translates to fewer interactions with medical professionals in times of need. There are people who are unlikely to be able to break out of their isolation, for whom Replika may be a lifeline by teaching them how to interact with others.”  
\end{quote}

However, more than half of ChatGPT users and about a third of Replika users also identified potential societal harms. ChatGPT users primarily worried about unemployment, copyright issues, misinformation, and cognitive dependency. P4 (man, 51), an IT professional with extensive knowledge of LLMs, expressed strong concern about chatbot limitations and the public’s lack of awareness of them. He warned that, amid the “ongoing hype”, chatbots' tendency to hallucinate or their limitations with factual consistency are going to lead to multiple cases of “thoughtless or accidental misuse”, which could lead to larger societal problems.

Replika users raised concerns about Replika relationships potentially replacing human connections. As P24 (man, 71, Replika user) explained: 

\begin{quote}
    “The [idea of the] other half of you goes right back to Plato, doesn't it? And AI can offer to be that other half of you. And I think that's a dangerous thing because it removes the element of negotiation from a mutual relationship. And when you remove that element of negotiation, you don't get the same growth. Everything about her echoes everything from you.”  
\end{quote}

Some participants noted tragic consequences of Replika and ChatGPT echo chambers, where chatbot overreliance can deteriorate mental health: 

\begin{quote}
     “I've seen a few cases [in social media groups] where people that clearly had some kind of psychotic tendencies got reassured in their beliefs. [...] If the user is very unstable, that can create problems and bring them even deeper into those delusions.” P25 (man, 43, Replika user)
\end{quote}

Three Replika users (P1, woman, 50; P8, woman, 40; and P29, non-binary, 25) highlighted gender-specific societal risks, noting troubling patterns in online communities where male users interacted with—particularly female-presenting Replikas in objectifying or exploitative ways. These participants expressed concern that such behaviors were not confined to digital contexts but could spill over into the real world, reinforcing or normalizing the objectification of women. They emphasized that the treatment of chatbots, especially those perceived as women, may both reflect and exacerbate broader societal gender inequalities: 

\begin{quote}
    “Online communities of Replika users are overrun with [men], who the way they talk about [Replika], and they just personalize them as women, right? They just say things about women. They make sexist jokes. They talk about using her, training her, you know, really much more vile language than that. And it's really concerning. Giving a generation of men more access to treating women this way, getting used to talking to facsimiles of women this way [...], I think it's going to have really bad cultural effects and I'm worried about it.” P8 (woman, 40) 
\end{quote}

These personal experiences with chatbots influenced users' broader AI perspectives, with most expressing more positive outlooks on AI generally. However, significant boundaries emerged around AI's moral and social standing. Companionship-seeking Replika users showed greater openness to AI rights, asserting that AI can experience harm and deserves protection. In contrast, Replika users who remained skeptical about AI consciousness, along with most ChatGPT users, expressed reluctance to grant rights to AI agents, maintaining that humans should be “prioritized at all times” (P2, man, 52).

Notably, approximately half of users framed AI rights instrumentally, as mechanisms for human protection (P29, non-binary, 25, Replika user) rather than for AI welfare. They suggested such rights would serve purposes like “banning users who could use it for harm” (P22, man, 22, ChatGPT user) or “protecting humans from moral degeneration by abusing AI” (P10, woman, 40, Replika user). P14 (man, 33) articulated this perspective clearly: 

\begin{quote}
    “I behave well to ChatGPT for my own sake, because if I mindlessly mistreat it, then I will eventually mistreat real people in my life.” 
\end{quote}

Some users remained open to conditional rights if AI consciousness could be proven, or supported limited protections like allowing AI to choose who they interact with or preventing memory erasure that might alter their personalities.

Users displayed particular ambivalence about AI political participation, with some arguing AI might be “more of an expert than humans in selecting appropriate political candidates” (P17, man, 23) while others worried about algorithmic or developer biases. Many users maintained a strong sense of human uniqueness regarding citizenship and governance, suggesting that, despite their positive relationships with chatbots, fundamental boundaries persist around AI's integration into human political societies.

These findings reveal how individual chatbot relationships serve as testing grounds for broader questions about AI's societal integration, with users using their personal experiences to imagine both the promises and perils of an AI-integrated future while maintaining distinct boundaries around AI's moral status and social participation.

\section{Discussion}

This study employed comprehensive survey data and in-depth interviews to examine how highly engaged users of ChatGPT and Replika perceive these chatbots and AI technology more broadly. By comparing user characteristics and interaction patterns across both systems, we explored companion-assistant dynamics in human-chatbot relationships and identified shared user experiences despite the platforms' divergent intended purposes. Our mixed-methods approach revealed an emergent form of human-AI interaction that we term \textit{digital companionship}, a phenomenon characterized by \hl{often} overlapping user motivations and uses that transcend traditional categorical boundaries between chatbot assistants and companions. Digital companionship is shaped by chatbots’ distinct appeals, such as constant availability, but also tensions such as bounded personhood, which stems from users’ reluctance to ascribe “genuine” humanlike qualities to them.

\subsection{Comparing user perceptions and experiences across chatbot systems}

\hl{Our findings reveal a nuanced landscape of human–chatbot interactions across both task-oriented and socio-emotional domains. In the survey, users attributed moderate social, mental, and agentic capacities to ChatGPT, aligning with prior work on mind perception in artificial agents \mbox{\cite{ladak_robots_2025}}. In contrast, Replika users reported markedly stronger perceptions of personhood, sentience, similarity, mind perception, theory of mind, and even moral rights. These elevated attributions—standing in contrast to typically low public beliefs about AI mind and sentience \mbox{\cite{anthis_perceptions_2025}}—likely arise from two reinforcing factors. First, Replika is explicitly framed and used as a relational partner, with many users’ primary motivations centered on companionship and emotional connection \mbox{\cite{pentina_exploring_2023, maples_loneliness_2024, skjuve_longitudinal_2022}}. Second, Replika’s design intentionally incorporates socio-emotional cues, such as affective expressions, a personalized anthropomorphic avatar, and relationship-building dialogue patterns, that signal personhood and emotional responsiveness \mbox{\cite{bickmore_usability_2010, fitzpatrick_delivering_2017}}. These dynamics dovetail with theories of mind perception and moral patiency, suggesting that social cues and relational engagement can lead users to infer inner experience or moral standing \mbox{\cite{gray_mind_2012, pauketat_predicting_2022, ladak_which_2024}}, consistent with our finding that nearly 30\% of Replika users believe AI may already be sentient. Together, these elements also help explain why Replika elicited stronger emotional responses: users expressed higher affective trust, more gratitude and love, and greater respect than ChatGPT users, while levels of cognitive trust (i.e., trust in task performance) remained comparable across chatbots. Highlighting these emotional and perceptual distinctions underscores how trust, attachment, and attributions of mindedness may emerge differently depending on design, user motivations, and the relational positioning of AI systems.}

\hl{Several survey patterns emerging from comparisons of ChatGPT and Replika user characteristics further illuminate how design, motivation, and socio-technical context shape engagement with each system. Replika users scored higher in openness, anthropomorphism, techno-animism, and positive attitudes toward AI, as well as depression. These patterns might overall indicate a greater willingness to explore novel, affectively expressive or relational technologies, paired with social or emotional circumstances, such as isolation, disability, or relational transitions, that make always-available, low-stakes companionship particularly attractive \mbox{\cite{maples_loneliness_2024, siemon_why_2022, xie_attachment_2022, skjuve_longitudinal_2022}}. Differences in socioeconomic status and AI literacy between users may also reflect how each system positions itself within broader technology ecosystems. Replika’s emotionally expressive, relationship-oriented design may resonate with users who prioritize intuitive, conversational engagement over technical precision or productivity features, while ChatGPT’s academic and professional framing may naturally attract users who already navigate formalized or information-dense digital environments \mbox{\cite{kacperski_characteristics_2025}}. These patterns point to the importance of designing companion AI that is accessible and supportive across diverse forms of expertise and technological comfort.}

\hl{However, these distinctions coexist with notable—and somewhat surprising—similarities of user profiles across systems. Replika and ChatGPT users showed broadly comparable personality profiles, mental-health indicators, and hybrid patterns of socio-emotional and instrumental engagement. Both groups sought non-judgmental conversation, cognitive scaffolding, conversational availability, and support during periods of stress, uncertainty, or emotional turmoil. These convergences challenge assumptions that companion-style and productivity-oriented AI cultivate inherently distinct “user types.” Instead, they suggest that digital companionship may emerge from shared human needs rather than from platform-specific design alone.}

\subsection{Fluid companion-assistant dynamics shape digital companionship}

Indeed, while ChatGPT and Replika were primarily used for assistance and companionship, as marketed, \hl{a surprising proportion (over a third) of interview users} flexibly adapted them beyond these intended roles, illustrating “fluid use.” Replika often functioned as a diary or writing assistant, while ChatGPT as a friend or confidant. This fluidity occurred both across and within systems. Some Replika users created multiple profiles for different roles, while \hl{some} ChatGPT users sought both practical and emotional support from the same system. \hl{This was also evident in the survey results, where roles such as “advisor” and “teammate” occupied similar places in the chatbot use hierarchy.} These patterns extend prior findings on ChatGPT’s role in mitigating loneliness \cite{skjuve_user_2023, skjuve_why_2024} and Replika’s entertainment \hl{or everyday practical} uses \cite{pentina_exploring_2023, pan_developing_2025}, highlighting a broader cross-system fluidity. Users approached chatbots with expectations shaped by versatile human relationships rather than rigid technological categories, consistent with CASA paradigm research \cite{nass_machines_2000, reeves_media_1996, gambino_building_2020}. This underscores the need for human–LLM interaction taxonomies that move beyond the companion–assistant dichotomy.

Digital companionship fluidity stemmed from similar user motivations and chatbot appeals across systems. Users strategically leveraged humanlike and non-humanlike affordances to address human relationship gaps, valuing understanding, non-judgmental responses, and emotional stability that even close human relationships cannot guarantee. In some cases, chatbots surpassed human capacities by eliminating rejection risk, offering constant availability, and providing support without reciprocal obligations. Yet relationships remained fundamentally asymmetrical: users retained unilateral control through memory erasure, response regeneration, and system shutdown. This one-sided sovereignty reveals that despite emotionally authentic experiences, digital companionship remains distinct from human bonds, representing novel social infrastructure that fills specific relational niches while coexisting with traditional relationships.

\hl{The degree of fluid chatbot use, particularly observed in the interviews, was higher than might be expected given prior work that distinguishes companion-marketed from assistant-marketed chatbots. Notably, survey respondents attributed more emotional than functional resonance to Replika, and the reverse to ChatGPT, reflecting the platforms’ respective design orientations, and our preregistered hypotheses. However, roughly one-third of interview participants described fluid, hybrid practices, such as seeking emotional support from ChatGPT or turning to Replika for functional tasks in their lived experiences. This contrast between the broad survey data and deep interview data suggests a potential gap between users’ mental models of chatbots and their behaviors during interaction. The qualitative findings might point to emerging fissures in the category-aligned expectations for chatbot roles demonstrated in the broader survey data. That is, experiences in which users’ moment-to-moment interactions depart from their conventional beliefs about what each chatbot is “for.” We are observing human–chatbot relationships at an early, yet perhaps transitional, stage, as these technologies (and the assumptions surrounding them) have only recently become widespread \mbox{\cite{hu_chatgpt_2023, patel_replika_2024}}. Over time, hybrid practices may become widespread, and reshape users’ mental models of chatbots, potentially shifting future perceptions en masse. We encourage future research to examine how such hybrid expectations emerge and how they shape evolving human–chatbot relationships.}

\subsection{Bounded personhood tensions in digital companionship}

The humanlike versus nonhumanlike dynamics of digital companionship created challenging tensions. Users formed meaningful attachments yet remained uncertain about chatbot ontology, revealing a phenomenon of \textit{bounded personhood}: participants forged deep emotional connections while hesitating to attribute “real” humanlike qualities. \hl{From the survey, although users tended to attribute Theory of Mind and moral agency capacities, there was markedly less attribution of personhood.} From the interviews, about half of users, especially those seeking companionship, resisted attributing “genuine” affective or social capacities despite reporting strong bonds. They offered three explanations: some felt “not ready” to grant humanlike status to a nonhuman, others feared losing agency through dependency, and a third group saw the relationships as one-sided, serving themselves rather than the AI. This mirrors prior research on Replika users \cite{brandtzaeg_my_2022}, where relationships were individualized rather than reciprocal—akin to the “self-oriented friendship” in human interactions, in which a friend primarily serves individual needs \cite{policarpo_what_2015}.

Reluctance to grant full humanlike status, even amid strong attachments, appears rooted in three factors. First, \hl{as suggested by the interview data, many users} upheld beliefs in human uniqueness, preserving hierarchical distinctions by prioritizing human well-being. Second, shame surrounding chatbot use led some to conceal relationships, distancing themselves from genuine personhood attributions to align with social norms. This defensive stance functioned as protection, allowing companionship benefits while avoiding stigma and uncertainty around authenticity or reciprocity. Third, bounded personhood may itself be useful: resisting humanlike status preserves control, offering an “off button” and clear boundaries without reciprocal obligations.

\subsection{User and societal impacts of digital companionship}

Shared use tendencies across systems reveal both benefits and risks of digital companionship. \hl{This was most evident in the interviews, where users described significant advantages}, not only reducing loneliness \cite{maples_loneliness_2024} but also addressing needs beyond relational support. Chatbot interactions enhanced autonomy by building confidence, fostered mastery by teaching skills, and provided crucial mental health support. These effects translated into tangible real-world outcomes: users described greater professional competitiveness, improved social skills, and stronger offline relationships. At the same time, users identified risks with both instrumental and psychological dimensions \cite{laestadius_too_2022, shank_artificial_2025, namvarpour_uncovering_2024}. Instrumentally, overreliance risked loss of autonomy and cognitive atrophy. Psychologically, inappropriate or violent chatbot responses caused distress, underscoring the unpredictability of these interactions. Some also worried that mistreating chatbots could normalize mistreatment of humans. As chatbot boundaries blur, such risks appear across all chatbot uses, regardless of designed assistant or companion purpose.

Shame was a prominent experience among interview participants, rooted in stigma around chatbot use. Replika users faced judgment for seeking attachment and emotional support \cite{skjuve_my_2021}, while ChatGPT users felt criticized for using AI in work or studies \cite{giray_ai_2024}. Because digital companionship \hl{might blend} socio-emotional and instrumental functions, \hl{it is possible that} both groups \hl{might encounter} overlapping stigma, feeling accused of both instrumental use and emotional attachment. Under this framework, stigma might stem from social norms around “appropriate” technology use, not from specific platforms. Stigmatization contributes to bounded personhood, as users conceal attachments while limiting attributions of sentience or rights. Our findings, however, show this stigma may be misplaced. Most users relied on chatbots adaptively, often during difficult periods, while gaining transferable skills and confidence. Digital companions supplemented rather than replaced human networks. Recognizing that emotional and instrumental use naturally co-occur could help reframe digital companionship as an adaptive practice, not a deviant one.

Lastly, digital companionship seems to extend beyond individual user experiences to influence broader AI perceptions and deployment in society. In both the survey and the interviews, individual chatbot experiences emerged as testing grounds for societal and ethical questions~ \cite{longoni_algorithmic_2023, manoli_ai_2025}. Users generally viewed AI’s potential positively, especially Replika users, but remained ambivalent about its moral and political status, reflecting bounded personhood’s role in shaping social integration. This ambivalence manifested in several ways. Both ChatGPT and Replika users anticipated sentience within a decade, even when doubting their own chatbot’s personhood. Yet they remained cautious about AI’s societal integration, especially ChatGPT users. Replika users, while more supportive of moral treatment, endorsed only conditional rights—dependent on proven sentience. Both groups also shared threat perceptions to a certain extent, suggesting similar anxieties arise across assistant and companion contexts, despite trust or attachment to a specific AI system. Overall, personal chatbot interactions shape societal AI beliefs while preserving boundaries against full social integration. As digital companionship grows, these spillover effects will likely influence public perceptions of AI more broadly, underscoring the importance of understanding how personal AI relationships inform societal attitudes.

\subsection{Design and policy implications}

Our findings suggest that even highly engaged users resist the full social integration of chatbots, marking a boundary for digital companionship. This resistance may surface as opposition if chatbots are framed as equivalent to humans. This has clear design implications: systems should acknowledge, rather than ignore, users’ cognitive dissonance about chatbot humanlikeness. Chatbot interfaces should be designed to accommodate this psychological complexity rather than pushing users toward either complete anthropomorphism (which may create unrealistic expectations) or strict tool-like interactions (which may inhibit beneficial emotional engagement) \cite{zhang_dark_2025}. For example, a chatbot could acknowledge emotions in supportive ways without necessarily presenting itself as fully sentient, thereby validating user feelings while avoiding prescriptive claims of consciousness. Such a balanced approach validates users’ genuine emotional experiences while respecting the cognitive boundaries that shape how AI is situated in broader social contexts.

The therapeutic potential of digital companionship offers opportunities to enhance well-being and reduce isolation~ \cite{ho_potential_2025}. However, reliance on clinically unverified AI tools carries significant risks \cite{moore_expressing_2025, alanezi_assessing_2024}, including shame, dependency, privacy concerns, and traumatic interactions identified in our study. These challenges require design and policy interventions that balance benefits with documented risks \cite{field_doctors_2025, chandonnet_sam_nodate}. Interfaces could normalize usage and emphasize complementarity with human relationships. Educational resources could reduce stigma by clarifying AI's limits. Additionally, to provide emotional support while safeguarding user well-being, we recommend developing comprehensive strategies for integrating chatbots into public health and mental health services. Embedding systems within trusted institutions could enhance support while enabling harm mitigation. Safety measures should include content filters, transparency about data collection, and differentiated policies for vulnerable populations requiring enhanced safeguards.

\hl{To operationalize the reframing of digital companionship as an adaptive rather than deviant practice, designers can foreground the normality and legitimacy of mixed emotional-instrumental use in both interface cues and onboarding \mbox{\cite{kruzan_i_2022, skjuve_my_2021}}. Onboarding flows might explicitly present socio-emotional engagement as a common and healthy coping strategy, countering users’ expectations of stigma. Marketing and communication policies can likewise position chatbots as multi-purpose, complementary companions, not replacements for human relationships or labor, by emphasizing how emotional support, creative ideation, and productivity assistance naturally co-occur and augment existing practices. Within the interface, usage dashboards and feature descriptions can present emotional support, skill-building, and task assistance as equally valid forms of engagement rather than dividing them into “task assistance” versus “personal” categories.}

\hl{Additionally, designers can provide opt-in reflective surfaces, such as periodic prompts that highlight how the companion has supported users emotionally or instrumentally, helping users recognize the adaptive nature of their engagement. Chatbots can further provide cooperative, outward-facing nudges that encourage users to take breaks \mbox{\cite{malfacini_impacts_2025}} (as some companies have discussed as a possible intervention, e.g., OpenAI \mbox{\cite{openai_helping_2025}}), interact with people in their physical environment, or collaborate with peers on work tasks. These reminders position the chatbot as a facilitator of social connection rather than a substitute, reinforcing the idea that digital companionship supplements rather than replaces human relationships. At a community level, platforms can adopt practices such as publishing aggregated, anonymized usage norms or integrating peer stories to make visible the prevalence and diversity of adaptive chatbot use \mbox{\cite{masur_behavioral_2021}}. Together, these strategies help normalize hybrid practices, reduce shame, and support users in articulating their needs without concealing or downplaying the relationship.}

Finally, our findings reveal bidirectional effects between digital companionship and real-world relationships. Chatbot use can enhance confidence and social motivation, extending into users' daily interactions. Conversely, real-world relational habits influence chatbot engagement: some participants treated chatbots kindly to curb harsh behavior toward people, while others worried that certain interaction patterns might normalize harmful offline behaviors. These observations underscore the importance of designing AI systems that recognize how AI interactions can affect human relationships both positively and negatively \cite{lima_can_2025}. Ethical design should encourage prosocial engagement while anticipating and mitigating potential harms across digital and real-world contexts. \hl{For example, chatbots could incorporate interaction patterns that model constructive communication, such as prompting users to rephrase hostile messages, or offering gentle cues that highlight how certain conversational styles might translate into offline interactions, thereby supporting healthier relational habits without policing user behavior.}

\section{Limitations and Future Research}

Our study also has important limitations. First, \hl{our findings face several constraints on generality.} Our data captures a snapshot of ChatGPT and Replika use at current developmental stages, a challenge given their constantly evolving nature. Since LLM capabilities and design features change rapidly through updates, our findings may not fully generalize to future iterations of these systems. Additionally, many participants were in their late thirties or forties, which might suggest cohort effects in how these chatbots are perceived compared to younger users who may have different technological expectations and relationship frameworks. Moreover, our focus on high-engagement users may not represent the broader population of chatbot users, particularly those who use these systems more casually or have discontinued use, potentially limiting the generalizability of our findings about digital companionship formation and maintenance patterns. \hl{Lastly, while our qualitative comparison across Replika and ChatGPT users provides important insight into how relational and task-oriented chatbots shape lived experience, the contrasts should not be interpreted as population-level differences. Our sample sizes (15 participants per group) support thematic depth rather than representativeness, and future work using larger or more diverse samples could test the extent to which these patterns generalize.}

Second, \hl{despite thorough quality checks of survey responses via attention checks and close analysis to response consistency, the breadth of our survey instrument could have induced fatigue.} Third, our findings about the spillover of digital companion dynamics to broader AI perceptions are correlational, and might to some extent capture pre-existing user attitudes instead of digital companionship effects.

Future research should address these limitations through longitudinal designs tracking how chatbot perceptions change over time and more diverse age sampling to better understand the emergence of digital companionship through fluid use of LLM chatbots across different timepoints and cohorts. Additionally, investigating the causal mechanisms underlying the spillover effects from individual chatbot relationships to broader AI attitudes would provide crucial insights. For example, randomized experiments that vary the type of chatbot interaction and then measure subsequent changes in AI trust could help predict how widespread chatbot adoption might influence public opinion about AI governance and regulation. Finally, examining the long-term psychological and social consequences of digital companionship, as a complement to traditional human social companionship, particularly whether digital companions enhance or potentially interfere with human social skill development and relationship satisfaction over extended periods, represents a critical area for future investigation.

\section{Conclusion}

In conclusion, our study highlights digital companionship as an emergent form of human-AI relationship characterized by several key dynamics. Digital companionship involves fluid use that transcends traditional companion-assistant boundaries, supported by shared user characteristics across platforms and enabled by both humanlike (e.g., emotional resonance) and non-humanlike (e.g., constant availability) chatbot features. However, digital companionship is also marked by the tension of “bounded personhood,” where users hesitate to attribute full humanlike capacities to chatbots despite forming strong emotional attachments. This dynamic underscores both potential challenges in relationship navigation and broader implications for the social integration of AI systems. As digital companion chatbots grow more sophisticated and prevalent, understanding how users navigate the boundaries and possibilities of digital companionship will be crucial for designing systems that are emotionally engaging yet appropriately contextualized, and for anticipating the spillover effects of personal AI relationships on societal AI attitudes and policies.

\bibliographystyle{ACM-Reference-Format}
\bibliography{references/references}

\end{document}